\documentclass[review]{elsarticle} % [preprint,8pt]{elsarticle}
\usepackage{graphicx}% Include figure files
\usepackage{xcolor}
\usepackage{adjustbox}
\usepackage{chemformula}
\usepackage{multirow}
\usepackage{dcolumn}% Align table columns on decimal point
\usepackage{bm}% bold math
\usepackage[mathlines]{lineno}% Enable numbering of text and display math
\usepackage[normalem]{ulem}
\usepackage{amsmath}
\usepackage{amssymb}
\usepackage{xcolor}
\usepackage[margin=1.0in]{geometry}

\journal{} % {Journal of Computational Physics}

\begin{document}
\begin{frontmatter}

\title{Linear stability analysis of compressible boundary layer over an insulated wall: Existence of multiple new unstable modes for Mach number beyond 3} %Title of paper

\author{Neha Chaturvedi$^{1}$} 
\ead{nehachaturvedi.19dr0071@mech.iitism.ac.in, chaturvedineha077@gmail.com}
\author{Swagata Bhaumik$^{2}$\corref{cor1}} 
\ead{swagatabhaumik@mech.iitism.ac.in}
\author{Rituparn Somvanshi$^{3}$} 
\ead{rituparn.19dr0126@mech.iitism.ac.in}
\cortext[cor1]{Corresponding author \\\it{Email address}: swagatabhaumik@iitism.ac.in}
\address{$^{1,2,3}${Department Of Mechanical Engineering, IIT (ISM)
    Dhanbad, Dhanbad - 826004, India}}

\begin{abstract}

We investigate the linear stability of parallel two-dimensional ($2D$) compressible boundary layer flow over a smooth, adiabatic flat plate. We consider both two- and three-dimensional ($3D$) disturbances, which yield $6^{th}$- and $8^{th}$-order systems for wall-normal variation of spectral amplitudes, respectively. We perform spatial stability analysis of the flow using the compound matrix method (CMM) to remove the stiffness of the problem, unlike the conventional techniques using Gram-Schmidt ortho-normalization or discretizing the governing equations by appropriate finite difference schemes. This method has yet to be employed for linear stability analysis of compressible boundary layers. We consider the Mach number $M$ of the flow to vary from low subsonic cases ($M=0.1$) to supersonic cases ($M=6$). First, we validate the compound matrix method by comparing the $M=0.1$ case results with the incompressible boundary layer governed by the Orr-Sommerfeld equation. The results depict an excellent match for primary and secondary modes. The flow becomes increasingly stable with an increase in $M$ up to $M=2$. We further note subsonic cases to be more stable when considering $3D$ disturbances. On the contrary, this conclusion holds only for $3D$ disturbances having spanwise wavenumber $\beta$ greater than a specific optimum value corresponding to supersonic cases up to $M=2$. We show that $2D$ disturbances are more spatially unstable than $3D$ perturbations with no spanwise variation (\textit{i.e.}, $\beta=0$) up to $M=1.4$, and the opposite scenario happens for $1.4<M<2$. Only one primary spatially unstable mode is found for boundary layer cases up to $M=2$. We locate a series of unstable modes for $M>3$, and the number of such modes is much more than two, as reported in Mack (AGARD Report No. 709, Part 3, 1984) from viscous calculations. Mack reports only two unstable modes up to $M=4.5$ that subsequently fuse with an increase in $M$. Our results show the number and the frequency extent of the unstable zone for these modes increase significantly with an increase in Mach number, \textit{e.g.}, we find $6$, $8$, and $10$ spatially unstable modes up to streamwise wavenumber of $\alpha_r=0.4$ for $M=3$, $4$ and $6$, respectively. The calculation of group velocity for these modes shows that these propagate downstream at a higher speed of $0.78$ to $0.89$ times the free-stream speed than those corresponding to incompressible, subsonic, and low supersonic ($M<2$) cases. While the shape of the neutral curves for the second unstable mode for $M=4$ and $6$ is similar to the fused neutral curve shown in Mack for $M=4.8$, the characteristics of spatially unstable higher-order modes, to the best of our knowledge, are not been reported elsewhere so far regarding the viscous stability of supersonic boundary layer.    
  
\end{abstract}


\end{frontmatter}

\section{Introduction} \label{sec_intro}

The stability of high-speed boundary layer and associated transition prediction is of great practical importance in designing subsonic and supersonic aircraft, hypersonic re-entry vehicles, gas turbine blades, and vanes \cite{xu2022,lee2019}. This information is essential to predict the above objects' skin friction, overall drag, and heat transfer characteristics. Significant theoretical \cite{mack1984,reed1996,dettenrieder2022}, experimental \cite{bitter2015,schneider2001}, and numerical \cite{malik1990,kosinov2015,fedorov2011,zhong2012,unnikrishnan2019} advances have been made to illustrate the transition to turbulence for high speed flows over the last several decades. A comprehensive account of various routes in this regard is given in Reshotko \cite{reshotko2008} and Lee \cite{lee2019}.

The transition of boundary layer flows from laminar to the turbulent stage is conventionally attributed to the growth of instability waves describing the evolution of small disturbances over the base flow \cite{tollmien1931,drazin2004}. Initial pioneering efforts to determine the stability of the incompressible boundary layer are due to Tollmein \cite{tollmien1931,tollmien1936}, Schlichting \cite{schlichting1933}, and Heisenberg \cite{heisenberg1924}, who obtained the analytical solution of the classical Orr-Sommerfeld equation (OSE) that governs the linear evolution of small perturbations in viscous incompressible flow framework, for a base flow that does not vary along streamwise direction (parallel flow approximation). These initial results demonstrated that complex interactions of viscous terms with the inertial terms of the governing equation can result in the amplification of monochromatic waves for flows that are stable according to the inviscid instability theories of Kelvin and Rayleigh \cite{kelvin1871,healey2006}. The growth of these viscous tuned monochromatic waves (later termed as Tollmien-Schlichting or $TS$-waves) in space/time is subsequently thought to be responsible for flow transition from laminar to turbulent state, ever since Schubauer \& Skramstad \cite{schubauer1947} reported corresponding experimental detections under controlled environment. 

The boundary layer transition process is more complicated for compressible flows or flows with heat transfer than the incompressible flows \cite{fedorov2011,zhong2012}. Spatial or temporal linear stability analysis of transonic and supersonic flows has been reported in Mack \cite{mack1975,mack1984} and Reed \cite{reed1996}. These studies are obtained by normal mode analysis in the form of temporally or spatially unstable disturbances. Mack \cite{mack1984} noted the existence of two unstable modes from $M=3$ to approximately $4.5$ for the viscous stability calculations of compressible flows over an insulated wall. The second mode corresponds to high wavenumber disturbances; therefore, its neutral curve is observed to lie on top of the first mode. Two neutral curves are noted to merge with an increase in Mach number to $4.8$ and beyond, denoting effectively the existence of one unstable zone whose spectral range is much wider than those corresponding to subsonic or low supersonic cases \cite{mack1984pof}. The existence of two separate unstable modes observed at intermediate supersonic Mach numbers is also noted for high-speed flows over hot and cold isothermal wall cases \cite{gasperas1989}, which subsequently, with an increase in flow Mach number, fuses to form a much wider single unstable zone. Experimental investigations on high-speed boundary layer transition confirm the existence and dominance of high-frequency (or high wavenumber) second mode \cite{kendall1975,demetriades1974,stetson1983,stetson1992}. Further details are also given in Fedorov \& Tumin \cite{fedorov2003} and Fedorov \textit{et. al.} \cite{fedorov2003JFM}.        

Conventionally, the stability calculations are performed by two approaches once the governing ordinary differential equation (ODE) for the spectral perturbation amplitudes under parallel flow approximation is obtained. The problem is cast as essentially an eigenvalue problem in the first approach, corresponding to the matrix obtained by discretizing the governing equations along the wall-normal directions by suitable numerical finite-difference schemes or using Chebyshev polynomials \cite{bertolotti1991,malik1982,malik1990}. This method is more appropriate for determining the temporal instability of the boundary layer as the streamwise and spanwise wavenumbers appear nonlinearly in these equations. Moreover, most of the eigenvalues provided by this method are numerical and spurious, whose number depends upon the number of discretization points and equations involved. Screening out the few real physical eigenmodes selectively from this pool of spurious ones is challenging. An alternative second approach for stability calculations of the viscous boundary layer is illustrated in Mack\cite{mack1976,mack1984}, which does not yield numerous spurious modes and is suitable for analysis of spatial, temporal, or even spatiotemporal instability of the boundary layer. Following this method, integration of spectral amplitude equations by suitable numerical schemes like Runge-Kutta or Runge-Kutta-Fehlberg methods \cite{gasperas1989,ozgen2008} is carried out from the known initial conditions at the free-stream. Straightforward numerical integration is impossible as the resulting coupled ODEs are inherently stiff, amplifying the spurious modes. Therefore, Gram-Schmidt ortho-normalization is applied at every step to circumvent this issue\cite{mack1976,gasperas1989,ozgen2008}. Subsequently, the eigenvalues are obtained by satisfying the appropriate dispersion relation at the wall.  

Ng \& Reid \cite{ng1985}, Yiantsios \& Higgins \cite{yiantsios1988} and Allen \& Bridges \cite{allen2002} have proposed the \textit{compound matrix method} (CMM) to remove the associated characteristic stiffness of such ODEs so that one can perform straightforward numerical integration without the solution diverging due to the appearance of growing non-physical modes. Compared to the method proposed by Mack\cite{mack1976}, here, no orthonormalization is required; instead, the integration is carried over the compound variables, which are essentially the minors of the modes representing the physically realizable solutions. This method has been successfully applied for the stability calculations of incompressible wall-bounded \cite{drazin2004,sengupta2012} and mixed-convection boundary layer \cite{sengupta2013} problems. CMM has also found applications in solving stiff governing equations related to the stability of plane Poiseuille flow containing multiple intermediate interfaces \cite{anturkar1990}, nonlinear traveling waves \cite{gubernov2006}, and pre-stressed elastic tube under axial compression \cite{haughton2008}. This method has so far not been used for stability calculations of the compressible boundary layers for which the constitution of corresponding compound matrix equations are more complex than incompressible hydrodynamic or mixed-convection boundary layer, as illustrated later in Sec.~\ref{sec_IIB}. It is performed here.

Results show that while the instability characteristic matches existing results for subsonic and low supersonic cases, we report the existence of multiple unstable modes for $M>3$ with increasing wavenumber. The number of such unstable modes increases with an increase in Mach number beyond $3$. In contrast, Mack \cite{mack1984} or subsequent similar investigations \cite{gasperas1989,ozgen2008} report at most two distinct unstable modes at intermediate supersonic Mach number cases. Our calculations show that the extent of the corresponding unstable spectral range is much broader than subsonic or low supersonic cases, which is enhanced with an increase in $M$. Several such unstable modes are characterized and illustrated in Sec.~\ref{sec_M_ge_3}. It is the novel element in the present manuscript. The existence of multiple (much more than two) unstable viscous modes has not been reported before for high-speed flows. 

The paper is organized as follows. Section~\ref{sec_linearstabilityanalysis} describes linearized disturbance equations, equations for respective spectral amplitudes and the far-field variation of disturbances corresponding to compressible boundary layers. Sections~\ref{sec_IIB} and \ref{sec_similarity} illustrate the dispersion function for determining the eigenmodes of compressible shear-layer over the insulated wall and wall-normal variation of the mean flow as obtained from corresponding self-similar equations\cite{stewartson1964}, respectively. Section~\ref{sec_results} provides the results of the instability analysis as obtained from CMM for the compressible boundary layer over the adiabatic wall. We start with validating the CMM methodology by comparing the results for $M=0.1$ with corresponding incompressible results in Sec.~\ref{sec_validation}. Subsequently, we explore the effects of variation of Mach number and spanwise wavenumber ($3D$-effect) on the flow instability in Secs.~\ref{sec_comp} and \ref{sec_beta_effect}, respectively. Results for high supersonic Mach numbers ($M=3$, $4$, and $6$) are provided in Sec.~\ref{sec_M_ge_3}, which describes the characteristics of multiple unstable modes for these flow cases. The summary and conclusion are provided in Sec.~\ref{sec_conclusion}.   

%%%%% ------------------------------------------------------------

\section{linearized disturbance equations for compressible boundary layer} \label{sec_linearstabilityanalysis}

The dimensional form of the governing unsteady Navier-Stokes equations (NSE) for the compressible flow, in vectorial notations are given as
\begin{eqnarray}
\frac{\partial \tilde{\rho}}{\partial \tilde{t}}+ \tilde{\nabla} \cdot \left(\tilde{\rho} \tilde{\bf v}\right) = 0 \label{eq1} \\ 
\frac{\partial }{\partial t} \left(\tilde{\rho} \tilde{\bf v} \right) + \tilde{\nabla} \cdot \left( \tilde{\rho} \tilde{\bf v}\tilde{\bf v}\right) =
- \tilde{\nabla} \tilde{p} + \tilde{\nabla} \cdot \tilde{\bf \tau} \label{eq2} \\
\frac{\partial }{\partial t} \left(\tilde{\rho} \tilde{e}_t \right) + \tilde{\nabla} \cdot \left( \tilde{\rho} \tilde{\bf v} \tilde{h}_t \right) =
\tilde{\nabla} \cdot \left( \tilde{\bf \tau} \cdot \tilde{\bf v} \right) - \tilde{\nabla} \cdot \tilde{\bf q} \label{eq3} 
\end{eqnarray}

\noindent where $\tilde{\rho}$ represents fluid density; $\tilde{T}$ represents fluid temperature; $p$ represents thermodynamic pressure; $\tilde{\bf v}$ represents fluid velocity; $\tilde{e}_i$ and $\tilde{e}_t=\frac{1}{2} \left( \tilde{\bf v}\cdot \tilde{\bf v} \right) + \tilde{e}_i$ define the specific internal and total energy, respectively;
$h_i=\tilde{e}_i+\tilde{p}/\tilde{\rho}$ and $\tilde{h}_t=\tilde{e}_t+\tilde{p}/\tilde{\rho}$ represent specific internal and total enthalpy, respectively; $\tilde{\bf q}$ represents
the heat-flux and $\tilde{\bf \tau}$ represents the viscous stress tensor. For an isotropic and Newtonian fluid obeying Fourier law of heat conduction, the symmetric
stress tensor and the heat-flux vector are given as $\tilde{\bf \tau} = -\tilde{\lambda} D {\bf I} + 2\tilde{\mu} \tilde{\bf \epsilon}$ and $\tilde{\bf q} = - \tilde{\kappa} \tilde{\nabla} \tilde{T}$,
respectively \cite{Kundu}. Here, $\tilde{\bf \epsilon}=\frac{1}{2}\left( \tilde{\nabla} \tilde{\bf v}+\tilde{\nabla}\tilde{\bf v}^T\right)$, 
and $D=\tilde{\nabla}\cdot \tilde{\bf v}$ are the symmetric strain-rate tensor and the volumetric dilatation rate, respectively. The variables, $\tilde{\mu}$, $\tilde{\lambda}$ and $\tilde{\kappa}$
denote the dynamic and the second coefficient of viscosity and the thermal conductivity of the fluid, respectively. According to Stokes' hypothesis, \cite{Kundu}
$\tilde{\lambda} = 2\tilde{\mu}/3$, which indicates that the bulk viscosity coefficient of the flow $k=0$. Here, we assume a calorically perfect gas, and therefore, the equation of state is
given as $\tilde{p}=R\;\tilde{\rho} \; \tilde{T}$ so that $\tilde{e}_i=c_v \tilde{T}$ and $h_i=c_p \tilde{T}$ where, $R$, $c_v=R/(\gamma-1)$, $c_p=\gamma R/(\gamma-1)$, and $\gamma=c_p/c_v$
are the universal gas constant, specific heat at constant volume and constant pressure and ratio of specific heats, respectively. We also assume here that $\tilde{\mu}$, $\tilde{\lambda}$,
and $\tilde{\kappa}$ are functions of temperature $\tilde{T}$ alone.

We use the free-stream values for velocity $\tilde{U}_{\infty}$, temperature $\tilde{T}_{\infty}$, density $\tilde{\rho}_{\infty}$, pressure $\tilde{p}_{\infty}=R \tilde{\rho}_{\infty} \tilde{T}_{\infty}$, viscosity $\tilde{\mu}_{\infty}$, heat-conductivity $\tilde{\kappa}_{\infty}$ as the corresponding reference values to non-dimensionalize the above equations. The non-dimensional Reynolds number, 
Mach number and Prandtl number of the flow is given as $Re=\tilde{\rho}_{\infty} \tilde{U}_{\infty} L/\tilde{\mu}_{\infty}$, $M=\tilde{U}_{\infty}/\sqrt{\gamma R \tilde{T}_{\infty}}$ and
$Pr=c_p\mu_{\infty}/k_{\infty}$, respectively, where $L$ is the reference length-scale.       

\begin{figure}[!htbp]
    \centering
    \includegraphics[width=1.0\textwidth]{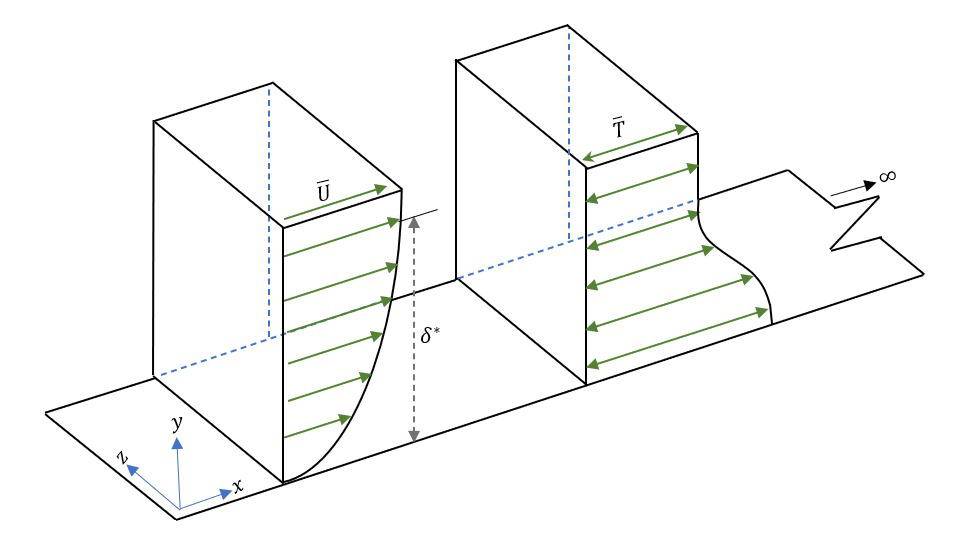}
    \caption{The schematic of the $2D$ parallel flow-approximation showing a tentative variation of the velocity and temperature profiles of a wall-bounded shear layer. Here, $\delta_*$ represents the
    displacement thickness of the shear-layer, which is treated as constant here under the parallel flow approximation.}
    \label{fig1}
\end{figure}

Next, to consider the disturbance evolution, we split each flow variable $\phi$ into a mean and fluctuating component as $\phi=\Bar{\phi}+\epsilon \hat{\phi}$, where $\epsilon<<1$. 
Therefore, we assume the fluctuating quantities are much smaller than the mean flow variables. The mean quantities are functions of spatial coordinates only, while the fluctuating quantities
depend on both space and time. By substituting the above decomposition into Eqs.~(\ref{eq1}-\ref{eq3}), we obtain the
linearized equations for the disturbance quantities. We denote $x$, $y$, and $z$ as the non-dimensional streamwise, wall-normal, and spanwise coordinates, respectively.
We consider a $2D$ parallel mean flow (as shown in Fig.~\ref{fig1}), and therefore, the mean quantities, \textit{i.e.}, $\bar{\varrho}$, $\bar{P}$, $\Bar{T}$ and $\Bar{U}$ are only functions of the
wall-normal coordinate $y$ only. Here, $\bar{U}$ is the streamwise component of the mean flow. As the boundary layer is assumed to be $2D$ and parallel, the wall-normal and spanwise component
of the mean flow is assumed to be zero, \textit{i.e.}, $\bar{V}=\bar{W}=0$. We consider the local displacement thickness $\delta_*$ as the
reference length-scale, \textit{i.e.}, $L=\delta_*$. Let, $\hat{\textbf{v}}=(\hat{u},\hat{v},\hat{w})$ represents perturbation velocity components, and $\hat{p}$, $\hat{\theta}$ and $\hat{\rho}$
denote disturbance pressure, temperature, and density, respectively. The non-dimensional linearized disturbance equations under $2D$ parallel flow approximation are given as \cite{mack1984,ozgen2008}

\begin{eqnarray}
  \frac{D\hat{\rho}}{Dt}+\hat{v}\frac{d\bar{\varrho}}{dy}+\bar{\varrho} \hat{D} & = & 0 \label{eq4} \\
  \bar{\varrho}\frac{D\hat{u}}{Dt} + \bar{\varrho}\hat{v} \frac{d\bar{U}}{dy} + \frac{1}{\gamma M^2} \frac{\partial \hat{p}}{\partial x} & = & \frac{1}{Re} \left( \frac{\partial \hat{\tau}_{xx}}{\partial x}+\frac{\partial \hat{\tau}_{xy}}{\partial y} +
  \frac{\partial \hat{\tau}_{xz}}{\partial z} \right) \label{eq5} \\
  \bar{\varrho}\frac{D\hat{v}}{Dt}+ \frac{1}{\gamma M^2} \frac{\partial \hat{p}}{\partial y} & = & \frac{1}{Re} \left( \frac{\partial \hat{\tau}_{yx}}{\partial x}+\frac{\partial \hat{\tau}_{yy}}{\partial y} + \frac{\partial \hat{\tau}_{yz}}{\partial y} \right) \\
  \bar{\varrho}\frac{D\hat{w}}{Dt}+ \frac{1}{\gamma M^2} \frac{\partial \hat{p}}{\partial z} & = & \frac{1}{Re} \left( \frac{\partial \hat{\tau}_{zx}}{\partial x}+\frac{\partial \hat{\tau}_{zy}}{\partial y} + \frac{\partial \hat{\tau}_{zz}}{\partial z} \right) \label{eq5a} \\
  \bar{\varrho} \bar{c}_p\frac{D\hat{\theta}}{Dt}-\frac{\gamma-1}{\gamma} \frac{D\hat{p}}{Dt} + \bar{\varrho} \bar{c}_p \hat{v} \frac{d\bar{T}}{dy} & = &
  \frac{1}{Re}\frac{\gamma-1}{M^2} \left[ \hat{\tau}_{xy}\frac{d\bar{U}}{dy} + \bar{\mu} \frac{d\bar{U}}{dy}\left(\frac{\partial \hat{u}}{\partial y} + \frac{\partial \hat{v}}{\partial x}\right) \right] \nonumber \\
  & & - \frac{1}{Re}\frac{1}{Pr} \left(\frac{\partial \hat{q}_x}{\partial x}+\frac{\partial \hat{q}_y}{\partial y}+\frac{\partial \hat{q}_z}{\partial z} \right) \label{eq6}  
%  & & {\color{red} \sout{\hat{u}\frac{d}{dy}\left(\bar{\mu}\frac{d\bar{U}}{dy}\right)}}-\left(\frac{\partial \hat{q}_x}{\partial x}+\frac{\partial \hat{q}_y}{\partial y}+\frac{\partial \hat{q}_z}{\partial z} \right)
\end{eqnarray}
\noindent where, the operator $\frac{D}{Dt}=\left(\frac{\partial}{\partial t}+\bar{U}(y)\frac{\partial}{\partial x}\right)$. The perturbation stress tensor components are given as 
$\hat{\tau}_{xx}=2\bar{\mu}\frac{\partial \hat{u}}{\partial x}+ \bar{\lambda}\hat{D}$,
$\hat{\tau}_{yx}=\hat{\tau}_{xy}=\bar{\mu}\left(\frac{\partial \hat{u}}{\partial y}+\frac{\partial \hat{v}}{\partial x}\right)+ \frac{d \bar{\mu}}{d\bar{T}} \frac{d\bar{U}}{dy}\hat{\theta}$,
$\hat{\tau}_{xz}=\hat{\tau}_{zx}=\bar{\mu}\left(\frac{\partial \hat{u}}{\partial z}+\frac{\partial \hat{w}}{\partial x}\right)$, 
$\hat{\tau}_{yy}=2\bar{\mu}\frac{\partial \hat{v}}{\partial y}+\bar{\lambda}\hat{D}$,
$\hat{\tau}_{yz}=\hat{\tau}_{zy}=\bar{\mu}\left(\frac{\partial \hat{w}}{\partial y}+\frac{\partial \hat{v}}{\partial z}\right)$, and
$\hat{\tau}_{zz}=2\bar{\mu}\frac{\partial \hat{w}}{\partial z}+\bar{\lambda}\hat{D}$ where $\hat{D}=\left(\frac{\partial \hat{u}}{\partial x}+\frac{\partial \hat{v}}{\partial y}+\frac{\partial \hat{w}}{\partial z}\right)$.
Components of perturbation heat-flux vector are given as $\hat{q}_x=-\bar{\kappa}\frac{\partial \hat{\theta}}{\partial x}$,
$\hat{q}_y= -\left( \bar{\kappa}\frac{\partial \hat{\theta}}{\partial y}+\frac{d \bar{\kappa}}{d\bar{T}} \frac{d\bar{T}}{dy}\hat{\theta} \right)$ and $\hat{q}_z=-\bar{\kappa}\frac{\partial \hat{\theta}}{\partial z}$. The relationship between $\hat{p}$, $\hat{\theta}$ and $\hat{\rho}$ are derived from the perturbation equation of state given as $\hat{p}=\left(\hat{\rho} \bar{T} +\bar{\varrho} \hat{\theta}\right)$.  

 \subsection{Variation of spectral amplitudes of the linearized disturbances} \label{spec} 

 Equations~(\ref{eq4}-\ref{eq6}) are subsequently expressed in the spectral domain by using the Fourier-Laplace transform of the perturbation quantities as
 
 \begin{eqnarray}
   \left(\hat{u},\hat{v},\hat{w},\hat{p},\hat{\rho},\hat{\theta}\right)^T=\int \int \left(\chi(y),\varphi(y),\Upsilon(y),\Pi(y),\zeta(y),\Theta(y) \right)^T
      e^{i\left(\alpha x +\beta z -\omega t \right)} d\alpha d\beta 
  \label{eq7} 
\end{eqnarray}

 \noindent where $\alpha$ and $\beta$ represent streamwise and spanwise wavenumber, respectively, while $\omega$ denote the circular frequency of the perturbation components. The
variables $\chi$, $\varphi$, $\Upsilon$, $\Pi$, $\zeta$, and $\Theta$ indicate the spectral amplitudes of disturbance quantities $\hat{u}$, $\hat{v}$, $\hat{w}$, $\hat{p}$, $\hat{\rho}$, and
$\hat{\theta}$, respectively. We define two auxiliary variables $\psi=\left(\alpha \chi + \beta \Upsilon\right)$ and $\Omega=\left(\beta\chi-\alpha\Upsilon\right)$. The variable $\psi$ represents
the spectral amplitude of $\hat{D_1}=\left(\frac{\partial \hat{u}}{\partial x}+\frac{\partial \hat{w}}{\partial z}\right)$, while $\Omega$ denotes the spectral amplitude of the perturbation
wall-normal component of vorticity $\hat{\xi}_y=\left(\frac{\partial \hat{u}}{\partial z}-\frac{\partial \hat{w}}{\partial z}\right)$. We note that in the context of the stability of
incompressible viscous flows, the Squire equation is essentially the wall-normal variation of the perturbation wall-normal vorticity component $\hat{\xi}_y$ \cite{schmid2002}.

Substituting the Fourier-Laplace transform of perturbations given by Eqs~(\ref{eq7}) into the linearized disturbance evolution equations~(\ref{eq4}-\ref{eq6}), we obtain the wall-normal variation of respective spectral amplitudes as \cite{mack1984,ozgen2008}

\begin{eqnarray}
iQ\zeta+\varphi\left(\frac{d\bar{\varrho}}{dy}\right)+\bar{\varrho}\left(\varphi'+i\psi\right) & = & 0 \label{appeq1} \\ 
%% Eqn-u
i\bar{\varrho}Q\chi+\varphi\left(\bar{\varrho}\frac{d\bar{U}}{dy}\right)+\frac{1}{\gamma M ^2} i\alpha\Pi & = & 
\frac{\bar{\mu}}{Re}\left[ \chi''-\Delta^2\chi+i\alpha \left(1+\frac{\bar{\lambda}}{\bar{\mu}}\right)\left(\varphi'+i\psi\right) \right] \nonumber \\
      & & +\frac{1}{Re} \biggl[ C_1 \left( i \alpha\varphi+ \chi' \right) +C _2 \Theta+C _3 \Theta' \biggl] \\ 
%% Eqn-v 
i\bar{\varrho}Q\varphi+\frac{1}{\gamma M ^2} \Pi' & = &
\frac{\bar{\mu}}{Re}\left[ \left(2+\frac{\bar{\lambda}}{\bar{\mu}}\right) \varphi''-\Delta^2\varphi+i\left(1+\frac{\bar{\lambda}}{\bar{\mu}}\right)\psi' \right] + \nonumber \\
 & & \frac{1}{Re}\biggl[i\frac{\bar{\lambda}}{\bar{\mu}}C_1 \psi+\left(2+\frac{\bar{\lambda}}{\bar{\mu}}\right)C_1 \varphi' + i\alpha C_3 \Theta \biggl] \\
% Eqn-w
i\bar{\varrho}Q\Upsilon +\frac{1}{\gamma M ^2} i\beta\Pi & = & %% + \varphi\left(\bar{\varrho}\frac{d \bar{W}}{dy}\right)
\frac{\bar{\mu}}{Re}\left[\Upsilon''-\Delta^2\Upsilon+i\beta \left(1+\frac{\bar{\lambda}}{\bar{\mu}}\right) \left(\varphi'+i\psi\right) \right] \nonumber \\
 & & + \frac{C_1}{Re} \biggl[ i \beta\varphi+ \Upsilon' \biggl] \\ % +C_5\Theta+C_4 \Theta' \right] \\
% Eqn-\theta
iQ\left(\bar{\varrho}c_p\Theta-\frac{\gamma-1}{\gamma}\Pi\right)+\left(\bar{\varrho}c_p\frac{d \bar{T}}{dy}\right)\varphi & = &
\frac{1}{Re}\frac{1}{Pr}\biggl[ \bar{\kappa}\left(\Theta''-\Delta^2\Theta\right)+C_9\Theta+2C _{10} \Theta' \biggl] + \nonumber \\
& & \frac{(\gamma-1)M ^2}{Re}\left[C_8\Theta+2\bar{\mu} \frac{d\bar{U}}{dy} \left(\chi'+i\alpha \varphi \right) \right] \label{appeq5} \\
% % \frac{d\bar{W}}{dy}\Upsilon' \right) + \nonumber \\
% & & 2i\bar{\mu}\left(\alpha \frac{d\bar{U}}{dy} + \beta \frac{d\bar{W}}{dy}\right) \varphi \biggl] \label{appeq5} % +C_7 \Upsilon\biggl] \label{appeq5}
%% Equation of state
\Pi=\left( \bar{T}\zeta+\bar{\varrho}\Theta \right)  \label{appeq6}
\end{eqnarray}

\noindent where, $(\cdot)^{\prime}=d(\cdot)/dy$, $\Delta=\sqrt{\alpha^2+\beta^2}$, and $Q=\left(\alpha\bar{U}-\omega \right)$. Other variables appearing in Eqs.~(\ref{appeq1}-\ref{appeq5}) are given as $C_1 = \frac{d{\mu}}{d \bar{T}} \frac{d \bar{T}}{dy}$,
$C_2 = \left(\frac{d^2 {\mu}}{d \bar{T}^2} \frac{d \bar{T}}{dy} \frac{d \bar{U}}{dy} + \frac{d{\mu}}{d\Bar{T}} \frac{d^2 \bar{U}}{dy^2}\right)$,
$C_3 = \left(\frac{d{\mu}}{d \bar{T}} \frac{d \bar{U}}{dy}\right)$, $C_8 = \left( \frac{d{\mu}}{d\bar{T}} \left( \frac{d \bar{U}}{dy}\right)^2 \right) $, %  \right)$,
$C_9 = \left( \frac{d^2 {\kappa}}{d \bar{T}^2} \left( \frac{d \bar{T}}{dy}\right)^2 + \frac{d{\kappa}}{d \bar{T}}\frac{d^2 \bar{T}}{dy^2} \right)$, and
$C_{10} = \left( \frac{d {\kappa}}{d \bar{T}} \frac{d \bar {T}}{dy}\right)$. Substituting Eq.~(\ref{appeq6}) in Eqs.~(\ref{appeq1}-\ref{appeq5}), and further
simplifying, one gets equations in terms of $\psi$, $\varphi$, $\Theta$ and $\Omega$ as 

\begin{eqnarray}
  % Eqn-\psi   
  \left[ \frac{\bar{\mu}}{Re} \right] \psi^{\prime\prime}
  + \left[\frac{C_1}{Re}\right] \psi^{\prime} 
  + \left[i\frac{\bar{\mu}}{Re}\left(1+\frac{\bar{\lambda}}{\bar{\mu}}\right) \Delta^2 +\frac{F_1}{\gamma M^2} \Delta^2 \right] \varphi^{\prime} % & \nonumber \\
  + \left[ \frac{\alpha C_3}{Re} \right] \Theta^{\prime}  & \nonumber \\ 
  + \left[ i\frac{F_1}{\gamma M^2}\Delta^2 - \frac{\bar{\mu}}{Re} \left(2+\frac{\bar{\lambda}}{\bar{\mu}}\right)\Delta^2 - i\bar{\varrho}Q \right] \psi % & \nonumber \\
  + \left[ \frac{F_1}{\gamma M^2}\frac{1}{\bar{\varrho}}\frac{d\bar{\varrho}}{dy}\Delta^2 - \alpha \bar{\varrho} \frac{d\bar{U}}{dy} +i\frac{C_1}{Re} \Delta^2 \right] \varphi
  & \nonumber \\  
  + \left[ \frac{\alpha C_2}{Re}-i\frac{\bar{\varrho}}{\gamma M^2} \Delta^2 \right] \Theta = 0 &
  \label{app1_eq1}
\end{eqnarray}
% Eqn-\varphi
 \begin{eqnarray}
  \left[ \frac{\bar{\mu}}{Re} \left(2+\frac{\bar{\lambda}}{\bar{\mu}}\right) Q^2 - i\frac{\bar{\varrho} \bar{T}}{\gamma M^2} Q \right] \varphi^{\prime\prime}  
  + \left[ i \frac{\bar{\mu}}{Re} \left(1+\frac{\bar{\lambda}}{\bar{\mu}}\right) Q^2 + \frac{\bar{\varrho} \bar{T}}{\gamma M^2} Q \right] \psi^{\prime} & \nonumber  \\
  + \left[ \frac{C_1}{Re} \left(2+\frac{\bar{\lambda}}{\bar{\mu}}\right) Q^2 - i \frac{G_2}{\gamma M^2} \right] \varphi^{\prime} 
  + \left[-\frac{\bar{\varrho}}{\gamma M^2} Q^2 \right] \Theta^{\prime}
  + \left[ i \frac{C_1}{Re} \left(\frac{\bar{\lambda}}{\bar{\mu}}\right) Q^2 + \frac{G_3}{\gamma M^2} \right] \psi & \nonumber \\
  + \left[ - \frac{\bar{\mu}}{Re} Q^2\Delta^2 - i \bar{\varrho} Q^3 - i \frac{G_4}{\gamma M^2} \right] \varphi
  + \left[ i \frac{\alpha C_3}{Re}-\frac{1}{\gamma M^2} \frac{d\bar{\varrho}}{dy} \right] \Theta = 0 & 
 \end{eqnarray}
 % Eqn-\Theta
 \begin{eqnarray}
  \left[ \frac{\bar{\kappa}}{Re Pr} \right] \Theta^{\prime\prime}  
  + \left[ 2 \alpha \frac{\bar{\mu}}{Re} \left(\gamma-1\right) \frac{M^2}{\Delta^2} \frac{d\bar{U}}{dy} \right] \psi^{\prime}
  + \left[-\frac{\gamma-1}{\gamma } F_1 Q \right] \varphi^{\prime} & \nonumber \\
  + \left[ 2 \frac{C_{10}}{Re Pr} \right] \Theta^{\prime}
  + \left[2 \beta \frac{\bar{\mu}}{Re}\left(\gamma-1\right)\frac{M^2}{\Delta^2} \frac{d\bar{U}}{dy} \right] \Omega^{\prime}
  + \left[-i\frac{\gamma-1}{\gamma } F_1 Q \right] \psi  & \nonumber \\  
  + \left[2i\alpha \frac{\bar{\mu}}{Re}\left(\gamma-1\right)M^2\frac{d\bar{U}}{dy}-\frac{\gamma-1}{\gamma }F_1\frac{Q}{\bar{\varrho}}\frac{d\bar{\varrho}}{dy}
    -\bar{\varrho}c_{p}\bar{T}\right] \varphi & \nonumber \\
  + \left[ -\frac{\bar{\kappa \Delta^2}}{Re Pr} + \frac{C_8}{Re} \left(\gamma-1\right)M^2 + \frac{C_9}{Re Pr}
    -i \bar{\varrho}Q\left(c_{p}-\frac{\gamma-1}{\gamma}\right)\right] \Theta = 0 &
  \label{app1_eq4} 
 \end{eqnarray}
 % Eqn-\Omega
 \begin{eqnarray}
  \left[ \frac{\bar{\mu}}{Re} \right] \Omega^{\prime\prime}  
  + \left[ \frac{\beta C_3}{Re} \right] \Theta^{\prime}
  + \left[ \frac{C_1}{Re}\right] \Omega^{\prime} % & \nonumber  \\
  + \left[-\beta \bar{\varrho} \frac{d\bar{U}}{dy}\right]\varphi+\left[ \frac{\beta C_2}{Re} \right] \Theta  
  + \left[ - \frac{\bar{\mu}}{Re} \Delta^2 - i\bar{\varrho}Q \right] \Omega = 0 &
  \label{app1_eq5}
 \end{eqnarray}

 \noindent where $F_1=\bar{\varrho}\bar{T}/Q$, $F_2=\frac{1}{\bar{\varrho}}\frac{d}{dy}\left(\frac{\bar{\varrho}^2\bar{T}}{Q}\right)$,
 $F_3=\frac{d}{dy}\left(\frac{\bar{T}}{Q}\frac{d\bar{\varrho}}{dy}\right)$, $G_2=F_2Q^2$, $G_3=\frac{dF_1}{dy}Q^2$, and $G_4=F_3Q^2$.
 
 For $3D$ disturbances ($\beta\ne 0$ and $\hat{w}\ne 0$), the above set of ODEs constitute an $8^{th}$-order system. However, when the
 disturbances are $2D$ in nature ($\beta=0$ and $\hat{w}=0$), it reduces to $6^{th}$-order system of ODEs \cite{mack1984}. For $2D$ disturbances, there will not be any equation for $\Omega$ as it is
trivially zero. This feature of Eqs.~(\ref{app1_eq1}-\ref{app1_eq5}), which governs the wall-normal
variation of the disturbance amplitudes in a parallel compressible boundary layer is different from its incompressible counterpart, \textit{i.e.}, the Orr-Sommerfeld equation (OSE) \cite{drazin2004}.
The OSE is a $4^{th}$-order ODE irrespective of whether the disturbances are $2D$ or $3D$.

For the stability calculations of the boundary layer on insulated wall, we need to satisfy perturbation no-slip, zero-normal velocity and zero-normal disturbance heat-flux conditions at the wall, \textit{i.e.}, $\hat{u}=\hat{v}=\hat{w}=\frac{\partial \hat{\theta}}{\partial y}=0$ at $y=0$. Hence, in terms of spectral amplitudes, we need to specify the following conditions at $y=0$, 

\begin{eqnarray}
  \psi(0)=\varphi(0)=\Omega(0)=0 \;\; \text{and} \;\; \Theta^{\prime}(0)=0
  \label{eq10a} 
\end{eqnarray}

\noindent These homogeneous wall conditions also need to be supplemented by decaying conditions on all perturbation variables in the free-stream, \textit{i.e.},
$\left(\psi(y),\chi(y),\varphi(y),\Upsilon(y),\Pi(y),\varrho(y),\Theta(y) \right)^T \rightarrow 0$ as $y\rightarrow \infty$. Next, we discuss the exponential decay rate of various modes in the free-stream.

\subsection{Far-field variation of disturbances} \label{far-field}

At the far-field ($y\rightarrow \infty$) all the disturbances should decay, \textit{i.e.}, $\left(\hat{\textbf{v}},\hat{\rho},\hat{\theta}\right)\rightarrow 0$ whereas $(\bar{U},\bar{\varrho},\bar{T},\bar{\mu},\bar{k})\rightarrow 1$. The Eqs.~(\ref{app1_eq1}-\ref{app1_eq5}), therefore, degenerate into a system of ODEs with constant coefficients. The exponential behavior of the modes is obtained from the corresponding characteristic polynomial equation. We can reduce the algebraic complexity by substituting $\varphi$, the spectral amplitude of the perturbation $v$-velocity, with $\Pi$, the spectral amplitude of disturbance pressure. Let, $\psi_{\infty}$, $\varphi_{\infty}$, $\Pi_{\infty}$, $\Omega_{\infty}$, and $\Theta_{\infty}$ denote the values of $\psi$, $\varphi$, $\Pi$, $\Theta$, and $\Omega$ at the free-stream (\textit{i.e.}, at $y\rightarrow \infty$). The degenerated system of ODEs at the free-stream in terms of $\psi_{\infty}$, $\Pi_{\infty}$, $\Theta_{\infty}$, and $\Omega_{\infty}$ are given as 

\begin{eqnarray}
  \psi^{\prime\prime}_{\infty} & = & \left[\Delta^2+iRe Q_{\infty}\right] \psi_{\infty} + b_{12}  \Pi_{\infty} + b_{13} \Theta_{\infty} \label{eq11} \\
  \Pi^{\prime\prime}_{\infty} & = & b_{22} \Pi_{\infty} + b_{23} \Theta_{\infty} \label{eq11a} \\
  \Theta^{\prime\prime}_{\infty} & = & b_{32} \Pi_{\infty} + b_{33} \Theta_{\infty} \label{eq11b} \\
  \Omega^{\prime\prime}_{\infty} & = & \left[\Delta^2+iRe Q_{\infty}\right] \Omega_{\infty} \label{eq11c}  
  \end{eqnarray} 

\noindent where, $b_{12} =\left[i\frac{Re}{\gamma M^2}-Q_{\infty}\left(1+\frac{\bar{\lambda}}{\bar{\mu}}\right)\right]\Delta^2$,
$b_{13} =\left[ Q_{\infty} \Delta^2 \left(1+\frac{\bar{\lambda}}{\bar{\mu}}\right) \right]$, $b_{22} = \tilde{b}_{22}/b_2$, $b_{23}=\tilde{b}_{23}/b_2$,
$b_2=\left[iQ_{\infty}\left(2+\frac{\bar{\lambda}}{\bar{\mu}}\right)+\frac{Re}{\gamma M^2} \right]$,
$\tilde{b}_{22} = \left[ \Delta^2 b_2 - Re Q^2_{\infty} \left( 1-\frac{\gamma-1}{\gamma}Pr \left(2+\frac{\bar{\lambda}}{\bar{\mu}}\right) \right) \right]$,
$\tilde{b}_{23}= Re Q^2_{\infty} \left[ 1-Pr \left(2+\frac{\bar{\lambda}}{\bar{\mu}}\right) \right]$, $b_{32} = -\left[i\frac{\gamma-1}{\gamma}Re Pr Q_{\infty}\right]$,
$b_{33} = \left[\Delta^2+iRe Pr Q_{\infty} \right]$, and $Q_{\infty}=\left(\alpha-\omega\right)$. Equations (\ref{eq11}-\ref{eq11c}) show that at the free-stream $\Omega_{\infty}$ is decoupled from $\varphi_{\infty}$, $\Pi_{\infty}$ and $\Theta_{\infty}$. Considering Eqs.~(\ref{eq11}-\ref{eq11c}), we note that modes vary in the free-stream as $\phi_j e^{-\left(\Lambda_j y\right)}$ where $j=1,\cdots,8$ and the exponents $\Lambda_j$ are given as   

\begin{eqnarray}
  \Lambda_{1,2} & = & \pm \sqrt{\left(\Delta^2+iRe Q_{\infty}\right)} \label{eq14ab} \\
  \Lambda_{3,4} & = & \pm \frac{1}{\sqrt{2}} \sqrt{\mathcal{T}_{\infty}+\sqrt{\mathcal{T}^2_{\infty}-4\mathcal{D}_{\infty}}} \label{eq14b} \\
  \Lambda_{5,6} & = & \pm \frac{1}{\sqrt{2}} \sqrt{\mathcal{T}_{\infty}-\sqrt{\mathcal{T}^2_{\infty}-4\mathcal{D}_{\infty}}} \label{eq14c} \\
  \Lambda_{7,8} & = & \pm \sqrt{\left(\Delta^2+iRe Q_{\infty}\right)} \label{eq14d}
\end{eqnarray}

\noindent where $\mathcal{T}_{\infty}=\left(b_{22}+b_{33}\right)$ and $\mathcal{D}_{\infty}=\left(b_{22}b_{33}-b_{23}b_{32}\right)$. Similar expressions of the behavior of the modes in the free-stream for compressible boundary layers are also given in Mack \cite{mack1984} and {\"O}zgen \& K{\i}rcal{\i} \cite{ozgen2008}, respectively. The free-stream variation of modes $j=1,2$ and $j=7,8$
are identical to that corresponding to the viscous mode obtained from the Orr-Sommerfeld equation\cite{sengupta2012,drazin2004}. Mack\cite{mack1984} postulated that as $Re\rightarrow \infty$, modes-$3$ and $4$ become independent of $M$ and depends only on $Re$ and $Pr$, while modes-$5$ and $6$ do not depend on $Re$ and $Pr$. Modes-$3$ and $4$, therefore, have been termed as the \textit{viscous-temperature mode}, while modes-$5$ and $6$ as the \textit{inviscid mode} in Mack\cite{mack1984}. This aspect is illustrated in the next section. As the physically realizable disturbances decay in the free-stream, we retain only the modes that decay in the free-stream. Hence, the physical eigenfunctions are linear combinations of modes $j=1,3,5$, and $7$ for $3D$ disturbances and $j=1,3$, and $5$ for $2D$ disturbances. In general, for $3D$ disturbances, therefore, we can denote $\Phi=\left(\psi,\varphi,\Theta,\Omega \right)^T$ as 

\begin{eqnarray}
  \Phi = c_1 \Phi_1 + c_3 \Phi_3 + c_5 \Phi_5 + c_7 \Phi_7
  \label{eq12a} 
\end{eqnarray}
\noindent whereas for $2D$ disturbances $\Phi$ can be expressed as
\begin{eqnarray}
  \Phi = c_1 \Phi_1 + c_3 \Phi_3 + c_5 \Phi_5 
  \label{eq12b} 
\end{eqnarray}

\noindent where $c_1$, $c_3$, $c_5$, and $c_7$ are arbitrary constants. Solving Eq.~(\ref{eq11}-\ref{eq11c}), the spectral amplitudes for the $j^{th}$ mode at the free-stream (where $j=1,3$ and $5$) are given as 
\begin{eqnarray}
     \begin{Bmatrix}
\psi_{j_{\infty}} \\
\varphi_{j_{\infty}}   \\
\Theta_{j_{\infty}} \\
\Omega_{j_{\infty}}
     \end{Bmatrix} = \begin{Bmatrix}
       b_{12}\left(b_{33}+\Lambda_j\right) -b_{13}b_{32} \\
       i\left[ Q_{\infty} \left(b_{32}+b_{33}+\Lambda_j^2\right) \left(b_{11}+\Lambda_j^2\right) - b_{12}\left(b_{33}+\Lambda_j^2\right)+b_{13}b_{32} \right]  \\
         {b_{32}\left(b_{11}+\Lambda_j^2\right)} \\
         0 \\
     \end{Bmatrix} e^{\left(-\Lambda_j y\right)}
     \label{eq12c}
  \end{eqnarray} 

\noindent where, $b_{11}=\left(\Delta^2+iRe Q_{\infty}\right)$. The free-stream variation for the Squire mode corresponding to $j=7$ (relevant only for $3D$ perturbations), is given as $\left[ \psi_{7_{\infty}}, \varphi_{7_{\infty}}, \Theta_{7_{\infty}}, \Omega_{7_{\infty}} \right]^{T}=\left[0,0,0,1\right]^{T} \; e^{\left(-\Lambda_7 y\right)}$.

\section{Formulation of the Compound Matrix Method (CMM) for calculation of eigenvalues} % \label{sec_formulation}
\label{sec_IIB}

\begin{figure}[!htbp]
    \centering
    \includegraphics[width=1.0\textwidth]{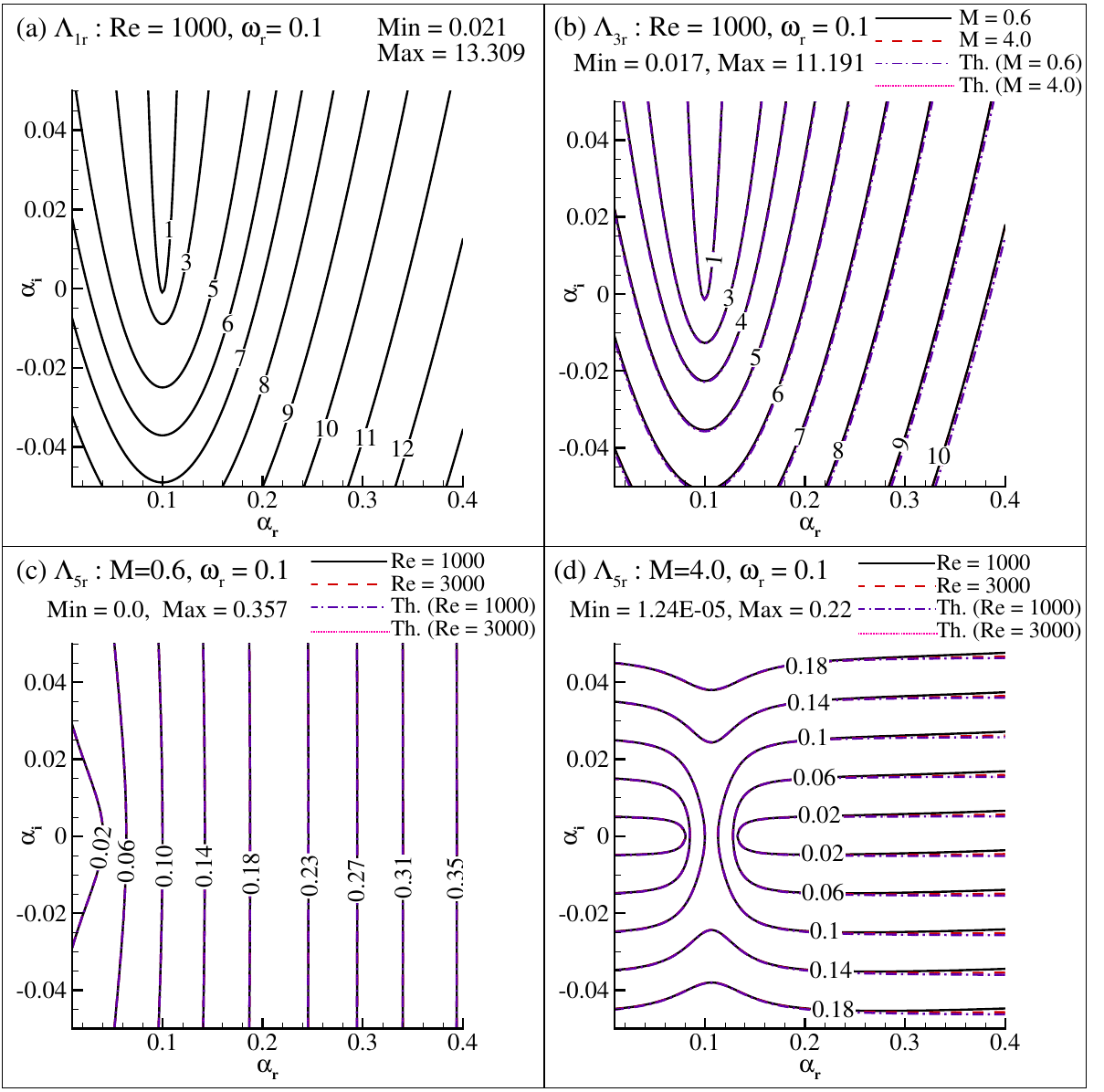}
    \caption{$\Lambda_{1_r}$, $\Lambda_{3_r}$, and $\Lambda_{5_r}$ plotted in the $(\alpha_r,\alpha_i)$-plane, where $\alpha=\alpha_r+i\alpha_i$, for indicated values of $Re$ and $M$ when $\omega_r=0.1$ and $\beta=0$.}
    \label{fig2}
\end{figure}

From the previous discussions, we note that physically realizable spectral amplitude of $3D$ disturbances that decay in the free-stream can be written as a linear combination of
four independent modes $\Phi_1$, $\Phi_3$, $\Phi_5$, and $\Phi_7$, respectively. These modes decay in the free-stream as $e^{-\Lambda_{1_r} y}$, $e^{-\Lambda_{3_r} y}$, $e^{-\Lambda_{5_r} y}$, and $e^{-\Lambda_{7_r} y}$,
respectively where the $\Lambda_{j_r}$ denotes the real part of $\Lambda_j$. In Fig.~\ref{fig2}, we plot contours of $\Lambda_{1_r}$, $\Lambda_{3_r}$, and $\Lambda_{5_r}$ in the
$(\alpha_r,\alpha_i)$-plane, ($\alpha=\alpha_r+i\alpha_i$) for indicated values of $Re$ and $M$ corresponding to $\omega_r=0.1$, $\omega_i=0$, and $\beta=0$, where $\omega=\omega_r+i\omega_i$.

Equation~(\ref{eq14ab}) suggests that $\Lambda_{1_r}$ is independent of the Mach number, and the corresponding contours are shown in Fig.~\ref{fig2}(a). We show the contours of $\Lambda_{3_r}$ for $M=0.6$ and $4$ for $Re=1000$ and $\omega_r=0.1$ in Fig.~\ref{fig2}(b). We note that contours of $\Lambda_{3_r}$ are virtually indistinguishable for $M=0.6$ and $4$. So, following Mack\cite{mack1984}, we identify this mode as the \textit{viscous-temperature} or entropic mode. Theoretically, for an ideal pure entropic mode, these exponents may be given as $\Lambda^{th}_{3,4}\approx \pm \sqrt{\left(\Delta^2+iRePr Q_{\infty}\right)}$. In Fig.~\ref{fig2}(b), we also show contours of $\Lambda^{th}_{3_r}$ for $M=0.6$ and $4$. We note that contour lines of actual $\Lambda_{3_r}$ and $\Lambda^{th}_{3_r}$ are almost identical, except at higher wavenumber regions, where slight deviation is noted. This observation justifies the postulation of Mack\cite{mack1984} in terming these modes as \textit{viscous-temperature} or entropic modes.       

Similarly, in Figs.~\ref{fig2}(c,d), we show the contours of $\Lambda_{5_r}$ for two different values of $Re$ for $M=0.6$ and $4$, respectively. Figure~\ref{fig2}(c) shows that for $M=0.6$, $\Lambda_{5_r}$ is practically independent of $\alpha_i$ when $\alpha_r>0.05$ as the contour lines are almost vertical. In contrast, for $M=4$, contours of $\Lambda_{5_r}$ is almost horizontal for $\alpha_r>0.15$ and therefore, $\Lambda_{5_r}$ is independent of $\alpha_r$ in this range. We also observe that as the Mach number is increased from $0.6$ to $4$, the values of $\Lambda_{5_r}$ reduce by one order of magnitude. For $M=4$, slight difference is noted between the contours for $\Lambda_{5_r}$ corresponding to $Re=1000$ and $3000$ for $\alpha_r>0.25$. Except this small difference, both Figs.~\ref{fig2}(c,d) show that contours of $\Lambda_{5_r}$ are virtually independent of $Re$. We denote $\Lambda^{th}_{5,6}=\pm \sqrt{\left(\Delta^2-Q_{\infty}^2M^2\right)}$, which is the exponent for the free-stream variation of the modes from inviscid instability analysis of compressible shear-layers\cite{mack1984}. We also show the contour lines of $\Lambda^{th}_{5_r}$ in Figs.~\ref{fig2}(c,d). We note that, while for $M=0.6$, contour lines of actual $\Lambda_{5_r}$ and $\Lambda^{th}_{5_r}$ are identical for both $Re$ cases, for $M=4$, slight deviation is noted at higher $\alpha_r$ regions. Almost perfect match of $\Lambda_{5_r}$ and $\Lambda^{th}_{5_r}$ in Figs.~\ref{fig2}(c,d) justifies the postulation of Mack\cite{mack1984} as these to be independent of $Re$ and $Pr$ when $Re\rightarrow \infty$. Therefore, we term this mode as the \textit{inviscid acoustic mode}.

As the modes decay at different rates, the straightforward integration of Eqs.~(\ref{app1_eq1}-\ref{app1_eq5}) from free-stream to wall is not possible. It would lead to the generation of spurious modes, which exponentially grow along the direction of integration, causing the numerical solutions to blow up. Ng and Reid \cite{ng1985} and Allen and Bridges \cite{allen2002} proposed the \textit{compound matrix method} (CMM) to circumvent this stiffness problem. In CMM, one solves a set of auxiliary equations derived from the original Eqs.~(\ref{app1_eq1}-\ref{app1_eq5}) in terms of compound variables. The compound variables are well-defined combinations of the fundamental modes $\Phi_j$ as defined in Eq.~(\ref{eq12a}) such that these auxiliary variables grow or decay exponentially at comparable rates. The application of the CMM removes the stiffness of the original perturbation equations, as also illustrated in Barker \textit{et. al.} \cite{barker2018} and Yiantsios \& Higgins \cite{yiantsios1988}.

Following this method, the $6^{th}$ order system for $2D$ disturbance would yield $^6C_3=20$ auxiliary compound matrix equations, while this number would be $^8C_4=70$ for $3D$ disturbances which follow the $8^{th}$ order stiff ODE. Using CMM, the original boundary value problem (BVP) is converted into an initial value problem (IVP), where the initial conditions are defined in the free-stream. The eigenvalues representing complex streamwise wavenumber $\alpha=\alpha_r+i\alpha_i$, for a particular combination of $Re$, $M$, $Pr$, $\omega$ and $\beta$ for spatial stability analysis, can be found by integrating these auxiliary equations from free stream to wall, subject to specified initial free-stream conditions, and satisfying the dispersion relation obtained from boundary condition at the wall as illustrated next.

The stability Eqs.~(\ref{app1_eq1}-\ref{app1_eq5}) can be recast as a system of first-order ODEs as
\begin{eqnarray}
  \left\{\textbf{X}^{\prime}\right\}=\left[E\right] \left\{\textbf{X}\right\} \label{eq15}
\end{eqnarray}
\noindent where the elements of the state vector $\left\{\textbf{X}\right\}=\left[X_1,X_2,X_3,X_4,X_5,X_6,X_7,X_8\right]^T$ are defined as $X_1=\psi$, $X_2=\psi^{\prime}$, $X_3=\varphi$, $X_4=\varphi^{\prime}$, $X_5=\Theta$, $X_6=\Theta^{\prime}$, $X_7=\Omega$ and $X_8=\Omega^{\prime}$. The matrix $\left[E\right]$ for $3D$ disturbances is a $8\times 8$ matrix. Similarly for the $2D$ case, $\left[E\right]$ is a $6\times 6$ matrix. Following Eq.~(\ref{eq12a}), we can express $\textbf{X}$ as as $\textbf{X} = c_1 \textbf{X}_1 + c_3 \textbf{X}_3 + c_5 \textbf{X}_5 + c_7 \textbf{X}_7$, where $\textbf{X}_1$, $\textbf{X}_3$, $\textbf{X}_5$ and $\textbf{X}_7$ are the linearly independent physical modes that decay in the free-stream.  

Following the methodology proposed in Allen and Bridges \cite{allen2002} for a fourth-order-system, we project the solutions of the above eighth-order system on a subspace of $C^8$ into $\bigwedge^4(C^8)$ with the help of $\textbf{X}_1$, $\textbf{X}_3$, $\textbf{X}_5$, and $\textbf{X}_7$. The problem is, therefore, reduced to linking these four-dimensional subspaces of $C^8$ satisfying Eqs.~(\ref{app1_eq1}-\ref{app1_eq5}) with a corresponding point in the vector space $\bigwedge^4(C^8)$. Mathematically, any subspace spanned by four linearly independent vectors $e_1$, $e_2$, $e_3$, and $e_4$ can be represented notationally as a point $e_1\wedge e_2 \wedge e_3 \wedge e_4$, in $\bigwedge^4(C^8)$. The boundary conditions at $y=0$ define a specific four-dimensional subspace of $C^8$, as illustrated later.

Let us define the new basis variables as $e_j=[X_{j,1},X_{j,3},X_{j,5},X_{j,7}]^T$, where $j=1,\cdots,8$ in $C^8$. Therefore, all elements of $e_j\wedge e_k \wedge e_l \wedge e_m$ form the basis for $\bigwedge^4(C^8)$ with the dimension $^8C_4=70$. The complete solution matrix, with these basis vectors, is obtained as

\begin{eqnarray}
  \begin{bmatrix}
X_{1,1} &  X_{1,3} & X_{1,5} & X_{1,7} \\ 
X_{2,1} &  X_{2,3} & X_{2,5} & X_{2,7} \\ 
X_{3,1} &  X_{3,3} & X_{3,5} & X_{3,7} \\ 
X_{4,1} &  X_{4,3} & X_{4,5} & X_{4,7} \\ 
X_{5,1} &  X_{5,3} & X_{5,5} & X_{5,7} \\ 
X_{6,1} &  X_{6,3} & X_{6,5} & X_{6,7} \\ 
X_{7,1} &  X_{7,3} & X_{7,5} & X_{7,7} \\ 
X_{8,1} &  X_{8,3} & X_{8,5} & X_{8,7} 
  \end{bmatrix}=
    \begin{bmatrix}
\psi_1            &  \psi_3            & \psi_5           & \psi_7 \\ 
\psi^{\prime}_1    &  \psi^{\prime}_3    & \psi^{\prime}_5    & \psi^{\prime}_7 \\ 
\varphi_1         &  \varphi_3         & \varphi_5         & \varphi_7 \\ 
\varphi^{\prime}_1 &  \varphi^{\prime}_3 & \varphi^{\prime}_5 & \varphi^{\prime}_7 \\ 
\Theta_1            &  \Theta_3            & \Theta_5           & \Theta_7 \\ 
\Theta^{\prime}_1    &  \Theta^{\prime}_3    & \Theta^{\prime}_5    & \Theta^{\prime}_7 \\ 
\Omega_1         &  \Omega_3         & \Omega_5         & \Omega_7 \\ 
\Omega^{\prime}_1 &  \Omega^{\prime}_3 & \Omega^{\prime}_5 & \Omega^{\prime}_7 
  \end{bmatrix}
  \label{eq16a}
\end{eqnarray}
\noindent Here, $X_{j,k}$ represents the $j^{th}$ element of the solution vector $\textbf{X}$ corresponding to $k^{th}$ mode, where $j=1,\cdots,8$ and $k=1,3,5,7$. Therefore, the seventy-two compound
variables are constructed as the $4\times 4$ minors of the solution matrix given by Eq.~(\ref{eq16a}). This can be symbolically denoted as

\begin{eqnarray}
  Y_n=\mathcal{Y}_{j,k,l,m}=
  \begin{vmatrix}
 X_{j,1} & X_{j,3} & X_{j,5} & X_{j,7} \\ 
 X_{k,1} & X_{k,3} & X_{k,5} & X_{k,7} \\ 
 X_{l,1} & X_{l,3} & X_{l,5} & X_{l,7} \\ 
 X_{m,1} & X_{m,3} & X_{m,5} & X_{m,7} 
\end{vmatrix}
  \label{eq16a}
\end{eqnarray}

\noindent where $1\le j < k < l < m \le 8 $ and $n=n(j,k,l,m)$. The details of the functional relationship between the indices are given in Appendix-$I$. Denoting $\left\{\textbf{Z}\right\}=\left[ Y_1,\cdots,Y_{70} \right]^T$, we get that $Z$ satisfies the set of linear coupled ODEs given by

\begin{eqnarray}
  \left\{\textbf{Z}^{\prime}\right\}=\left[F\right] \left\{\textbf{Z}\right\}
  \label{eq17}
\end{eqnarray}
\noindent Here, $[F]$ is $70\times 70$ matrix for $3D$ disturbance. We can determine the elements of the $[F]$ matrix from the elements of the matrix $[E]$ given in Eq.~(\ref{eq15}). The detailed methodology is illustrated in Appendix-$II$. One readily notes from Eq.~(\ref{eq16a}) that at the free-stream $Y_{n_\infty}\simeq \mathcal{C}_n \exp\left(-\left(\Lambda_1+\Lambda_3+\Lambda_5+\Lambda_7\right)y\right)$, where $\mathcal{C}_n$ is some constant. Therefore, all the modes in the free-stream decay at the identical exponential rate $e^{\left(-\left(\Lambda_1+\Lambda_3+\Lambda_5+\Lambda_7\right)y\right)}$, removing the stiffness of the original problem. Thus, we can solve Eq.~(\ref{eq17}) by using any standard integration procedure and no special treatment like repeated orthonormalization of the solution as adopted in Mack\cite{mack1976} and {\"O}zgen \& K{\i}rcal{\i} \cite{ozgen2008} is required. 

At the free-stream ($y\rightarrow \infty$) exact analytic nature of the modes $\Phi_1$, $\Phi_3$, $\Phi_5$ and $\Phi_7$ are known from Eq.~(\ref{eq12c}). This equation makes it possible to specify initial conditions for the compound variables $Y_n$ at the free-stream. Using the definitions of $Y_n$ given in Eq.~(\ref{eq16a}) where $n=1,\cdots,70$, and the analytical behavior of $\Phi_j$
(here, $j=1,3,5,7$) at the free-stream given by Eq.~(\ref{eq12c}), one obtains the initial conditions for $Y_n$ at $y \rightarrow \infty$ as
\begin{eqnarray}
  Y_{n_{\infty}}=\mathcal{Y}_{{j,k,l,m}_{\infty}}=
  \begin{vmatrix}
 X_{{j,1}_{\infty}} & X_{{j,3}_{\infty}} & X_{{j,5}_{\infty}} & X_{{j,7}_{\infty}} \\ 
 X_{{k,1}_{\infty}} & X_{{k,3}_{\infty}} & X_{{k,5}_{\infty}} & X_{{k,7}_{\infty}} \\ 
 X_{{l,1}_{\infty}} & X_{{l,3}_{\infty}} & X_{{l,5}_{\infty}} & X_{{l,7}_{\infty}} \\ 
 X_{{m,1}_{\infty}} & X_{{m,3}_{\infty}} & X_{{m,5}_{\infty}} & X_{{m,7}_{\infty}} 
\end{vmatrix} e^{\left(-\left(\Lambda_1+\Lambda_3+\Lambda_5+\Lambda_7\right)y_{\infty}\right)} 
  \label{eq17x}
\end{eqnarray}

\noindent The system of equation given by Eq.~(\ref{eq17}) now can be integrated as an initial value problem (IVP) from $y=y_{\infty}$ to wall \textit{i.e.}, $y=0$ by any standard ODE solving technique like forth-order Runge-Kutta method $RK_4$. While using $Y_{n_{\infty}}$ as the initial condition for Eq.~(\ref{eq17}) at $y=y_{\infty}$, it is preferred to scale the corresponding values by $e^{\left(-\left(\Lambda_1+\Lambda_3+\Lambda_5+\Lambda_7\right) y_{\infty} \right)}$. 

To find out the eigenvalues of Eqs.~(\ref{app1_eq1}-\ref{app1_eq5}), one needs to satisfy the homogeneous boundary conditions at $y=0$ given by Eq.~(\ref{eq10a}). From Eq.~(\ref{eq10a}), we can write at $y=0$
\begin{eqnarray}
  \psi(0) & = & c_1\psi_1(0)+c_3\psi_3(0)+c_5\psi_5(0)+c_7\psi_7(0) \nonumber \\
  & = & c_1X_{1,1}(0)+c_3X_{1,3}(0)+c_5X_{1,5}(0)+c_7X_{1,7}(0) = 0  \\
  \varphi(0) & = & c_1\varphi_1(0)+c_3\varphi_3(0)+c_5\varphi_5(0)+c_7\varphi_7(0) \nonumber \\
  & = & c_1X_{3,1}(0)+c_3X_{3,3}(0)+c_5X_{3,5}(0)+c_7X_{3,7}(0) = 0 \\
  \Theta^{\prime}(0) & = & c_1\Theta^{\prime}_1(0)+c_3\Theta^{\prime}_3(0)+c_5\Theta^{\prime}_5(0)+c_7\Theta^{\prime}_7(0) \nonumber \\
  & = & c_1X_{6,1}(0)+c_3X_{6,3}(0)+c_5X_{6,5}(0)+c_7X_{6,7}(0) = 0 \\
  \Omega(0) & = & c_1\Omega_1(0)+c_3\Omega_3(0)+c_5\Omega_5(0)+c_7\Omega_7(0) \nonumber \\
  & = & c_1X_{7,1}(0)+c_3X_{7,3}(0)+c_5X_{7,5}(0)+c_7X_{7,7}(0) = 0 \label{eq17a}
\end{eqnarray} 

\noindent To have a non-trivial solution, therefore, we need to satisfy the characteristic determinant of the linear system of equations to be identically zero at the wall, \textit{i.e.},

\begin{eqnarray}
  Y_{23}(0)=\mathcal{Y}_{1,3,6,7}(0)=
    \begin{vmatrix}
 \psi_1(0) & \psi_3(0) & \psi_5(0) & \psi_7(0) \\ 
 \varphi_1(0) & \varphi_3(0) & \varphi_5(0) & \varphi_7(0) \\ 
 \Theta^{\prime}_1(0) & \Theta^{\prime}_3(0) & \Theta^{\prime}_5(0) & \Theta^{\prime}_7(0) \\ 
 \Omega_1(0) & \Omega_3(0) & \Omega_5(0) & \Omega_7(0) 
\end{vmatrix} =  
    \begin{vmatrix}
 X_{1,1}(0) & X_{1,3}(0) & X_{1,5}(0) & X_{1,7}(0) \\ 
 X_{3,1}(0) & X_{3,3}(0) & X_{3,5}(0) & X_{3,7}(0) \\ 
 X_{6,1}(0) & X_{6,3}(0) & X_{6,5}(0) & X_{6,7}(0) \\ 
 X_{7,1}(0) & X_{7,3}(0) & X_{7,5}(0) & X_{7,7}(0) 
    \end{vmatrix}=0 \hspace{1cm} \label{eq17a}
 \end{eqnarray} 

\noindent The above equation provides the dispersion relation as $D_r +iD_i=Y_{23}(0)=\mathcal{Y}_{1,3,6,7}(0)=0$. Integrating the resultant auxiliary system of
Eqs.~(\ref{eq17}) subject to the initial conditions given corresponding to the normalized $Y_{n_{\infty}}$ at $y=y_{max}$, and satisfying the above dispersion relation, we can detect the
eigenvalues for a given set of parameters.
% , \textit{i.e.}, $Re$, $\bar{\lambda}/\bar{\mu}$, $\gamma$, $Pr$, and $M$.

For $2D$ disturbances, Eqs.~(\ref{app1_eq1}-\ref{app1_eq5}) constitute a $6^{th}$-order system, and hence, there would be three physical linear independent modes $\Phi_1$, $\Phi_3$, and $\Phi_5$ that decay in the free-stream. Consequently, the application of CMM would yield $^6C_3=20$ compound matrix equations. The rest of the procedure is similar to what is described above.

\section{Variation of mean wall-bounded shear layer over adiabatic flat plate} \label{sec_similarity}

\begin{figure}[!htbp]
\centering
\includegraphics[width=1.0\textwidth]{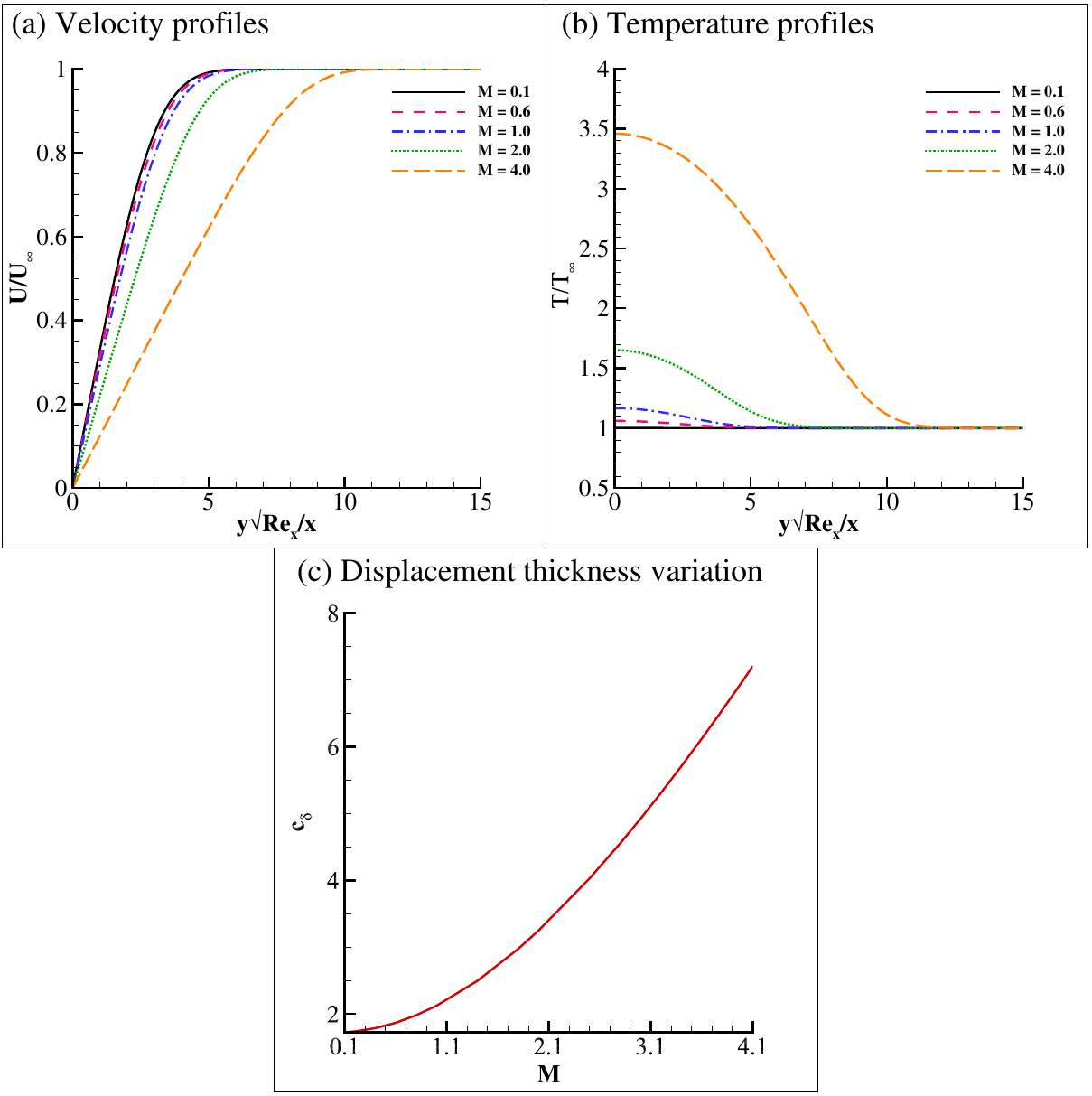}
    \caption{(a) $\bar{U}(y)$, and (b) $\bar{T}(y)$, plotted as a function of $\hat{\eta}=\tilde{y}\sqrt{Re_x}/\tilde{x}$, where $Re_x=\tilde{\rho}_{\infty}\tilde{U}_{\infty}\tilde{x}/\tilde{\mu}_{\infty}$ is the Reynolds number based on the streamwise coordinate $\tilde{x}$. (c) The displacement thickness parameter $c_{\delta}$ plotted as a function of the free-stream Mach number $M$ where, $c_{\delta}=\delta_*\sqrt{Re_x}/\tilde{x}$.}
    \label{fig3}
\end{figure}

For a $2D$ compressible wall-bounded shear layer, the self-similar equations can be obtained from the corresponding boundary-layer equations\cite{shapiro1954} following Illingworth \cite{white2006} or Howarth–Dorodnitsyn \cite{stewartson1964} transformation from the physical $(\tilde{x},\tilde{y})$-plane to transformed $(\xi,\eta)$-coordinate system, where $\xi=\int^{\tilde{x}}_0 \tilde{\rho}_e\tilde{\mu}_e \tilde{U}_e d\tilde{x}$, and $\eta=\frac{\tilde{U}_e}{\sqrt{\xi}}\int^{\tilde{y}}_0 \tilde{\rho} d\tilde{y}$. Here, $\tilde{\rho}_e$, $\tilde{U}_e$ and $\tilde{\mu}_e$ indicate density, streamwise velocity, and dynamic viscosity at the edge of the shear-layer. Here, we are not considering any external pressure gradient to be applied, and hence the edge conditions are not functions of streamwise coordinate and are treated to be constants, \textit{i.e.}, $\tilde{\rho}_e=\tilde{\rho}_{\infty}$, $\tilde{U}_e=\tilde{U}_{\infty}$, and $\tilde{\mu}_e=\tilde{\mu}_{\infty}$. The self-similar equations are given as \cite{stewartson1964}

\begin{eqnarray} \label{eq:similarityeqn1}   
\left(\bar{\varrho}\bar{\mu} f_{\eta\eta} \right)_{\eta}+ff_{\eta\eta} & = & 0 \\
    \left( \frac{1}{Pr} \bar{\varrho}\bar{\kappa} g_{\eta} \right)_{\eta} + \bar{c}_p fg_{\eta} + \left(\gamma-1\right)M^2 \bar{\varrho}\bar{\mu} f_{\eta\eta}^2 & = & 0
\end{eqnarray}

\noindent where $\left(\cdot\right)_{\eta}=d\left(\cdot\right)/d\eta$. The self-similar variables $f(\eta)$ and $g(\eta)$ are defined such that $f_{\eta}=\bar{U}$ and $g=\bar{T}$.
For no-slip adiabatic wall, one needs to satisfy $f(0)=f_{\eta}(0)=g_{\eta}(0)=0$ while at the free-stream $\left(f_{\eta},g\right)\rightarrow 1$ as $\eta\rightarrow \infty$. Following the
boundary layer approximation, $\frac{\partial \tilde{p}}{\partial \tilde{y}}=0$ in the shear layer, and hence, considering the ideal equation of state is obeyed, $\bar{\varrho}\bar{T}=1$. Here, we
consider the dynamic viscosity $\tilde{\mu}$, heat conductivity $\tilde{\kappa}$ and specific heat at constant pressure $\tilde{c}_p$ are functions of temperature only\cite{hilsenrath1955} as

\begin{eqnarray}
  \bar{\mu} & = & \bar{T}^{3/2}\left(\frac{1+110/T_{ref}}{\bar{T}+110/T_{ref}}\right) \label{eq14a} \\
  \bar{\kappa} & = & \bar{T}^{1/2} \left(\frac{1+S_3 10^{-S_4}}{1+\left(S_3/\bar{T}\right)10^{-S_4/\bar{T}}}\right) \\
  \bar{c}_p & = & \left( \frac{1+ \frac{\gamma-1}{\gamma} {\left(\theta_2^2e^{-\theta_2}\right)}/{\left(1-e^{\theta_2}\right)^2}}
      {1+ \frac{\gamma-1}{\gamma} {\left(\theta_1^2e^{-\theta_1}\right)}/{\left(1-e^{\theta_1}\right)^2}} \right)
  \end{eqnarray}

\noindent where $S_3=\left(245.4/T_{ref}\right)$, $S_4=12/T_{ref}$, $\theta_1=3055/T_{ref}$, $\theta_2=\theta_1/\bar{T}$ and $T_{ref}=303K$. The variation of dynamics viscosity denoted in
Eq.~(\ref{eq14a}) is the well-known Sutherland's law \cite{hilsenrath1955,white2006} for air. In Fig.~\ref{fig3}(a,b), we plot the self-similar velocity $\bar{U}$, and temperature $\bar{T}$ as a function of $\hat{\eta}=\tilde{y}\sqrt{Re_x}/\tilde{x}$ for the adiabatic plate condition. Here, $Re_x=\tilde{\rho}_{\infty}\tilde{U}_{\infty}\tilde{x}/\tilde{\mu}_{\infty}$ is the Reynolds number based on local streamwise coordinate $\tilde{x}$. We note that the increase in Mach number increases the hydrodynamic and thermal boundary layer thickness. As noted in Fig.~\ref{fig3}(b), this also increases the wall temperature. We use the displacement thickness $\delta_*$ as the reference length scale in the stability calculations to determine the Reynolds number $Re$. For any self-similar wall-bounded laminar shear layer, $\delta_*=c_{\delta}\tilde{x}/\sqrt{Re_x}$, where the factor $c_{\delta}$ is a function of Mach number $M$, $\gamma$ and Prandtl number $Pr$. For an incompressible, isothermal, and laminar flow, $c_{\delta}=1.72$ \cite{white2006}. In Fig.~\ref{fig3}(c), we show the variation of $c_{\delta}$ as a function of $M$. As the Mach number $M$ increases, $c_{\delta}$ monotonically increases.

\section{Results and discussion} \label{sec_results}

Traditionally, two approaches have been adopted to perform stability analysis of boundary layers, namely (1) temporal stability analysis, which considers complex $\omega$ with real $\alpha$ and $\beta$, and (2) spatial stability theory, which assumes complex $\alpha$ (and $\beta$) treating $\omega$ as real. A more elaborate and realistic analysis can be performed using spatiotemporal stability analysis where $\alpha$, $\beta$, and $\omega$ are all treated as complex. Detailed description in this regard may be found in Drazin and Reid \cite{drazin2004}, Tollmien \cite{tollmien1931}, Schmid \textit{et. al.} \cite{schmid2002}, and Sengupta \cite{sengupta2012}. Here, we only focus on the spatial stability analysis and treat $\omega$ and $\beta$ as real and $\alpha=\alpha_r+i\alpha_i$ as complex. Therefore, $\omega_i=0$, where $\omega=\omega_r+i\omega_i$. As illustrated before, spatial instability occurs when $\alpha_i<0$. 

\subsection{Comparison of spatial stability between incompressible and compressible boundary layers at $M=0.1$} \label{sec_validation}

\begin{figure}[!htbp]
    \centering
    \includegraphics[width=0.9\textwidth]{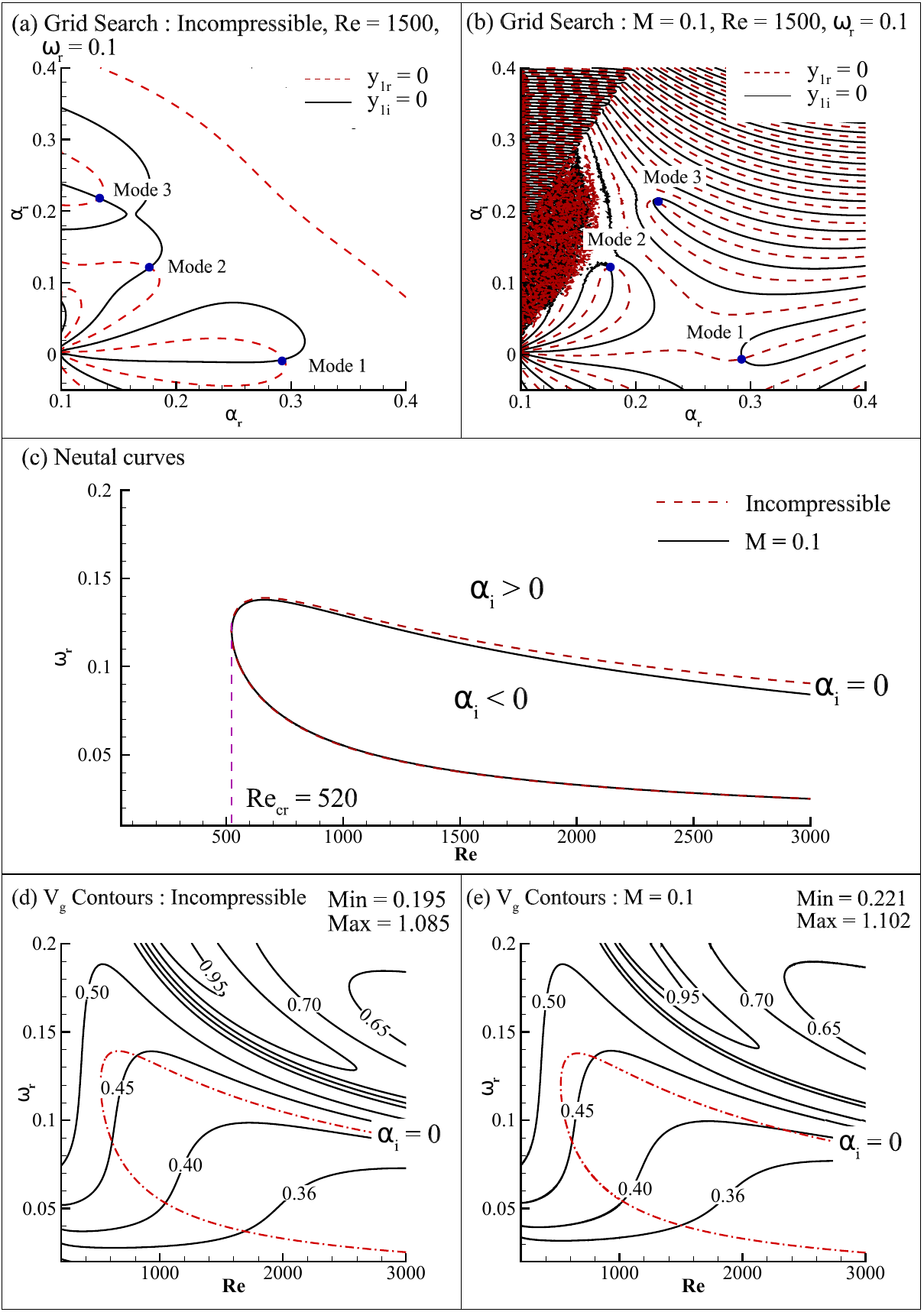}
    \vspace{-0.5cm} 
    \caption{Comparison between the spatial stability analysis of 2D incompressible zero-pressure gradient Blasius and compressible boundary layer on the insulated wall at $M=0.1$. (a,b) Spatial eigenvalues for $Re=1500$ and $\omega_r=0.1$, (c) comparison of the corresponding \textit{Neutral curves}, and (d,e) group velocity corresponding to mode-$1$ in the $(Re,\omega_r)$-plane.}
    \label{fig4}
\end{figure}

\begin{table}
\vspace{0mm}
\begin{center}
\begin{tabular}{| c | c | c | c | c |} %  c| c| c| c |}
\hline 
% \multicolumn{5}{c} \text{} \\             
\multicolumn{5}{c} \text{(a) CMM applied to OSE for 2D incompressible ZPG boundary layer}  \\ \hline              
Mode No. & $\alpha_r$ & $\alpha_i$       & $Y_{1r}(0)$          & $Y_{1i}(0)$ \\ \hline
%          &            &                  &                  &          \\  
 $1$  & $0.29373724$ & $-7.03994013\times 10^{-3}$ & $1.6941\times 10^{-19}$  & $6.8821\times 10^{-20}$   \\
 $2$  & $0.17675906$ & $0.12104521$               & $-2.9461\times 10^{-16}$ & $2.0643\times 10^{-17}$  \\
 $3$  & $0.13301162$ & $0.21791196$               & $-9.1898\times 10^{-14}$ & $2.4733\times 10^{-14}$  \\
 \hline
% \multicolumn{5}{c}  \text{} \\
\multicolumn{5}{c} \text{(b) CMM applied to 2D compressible boundary layer over adiabatic plate for $M=0.1$} \\ \hline
      Mode No. & $\alpha_r$  & $\alpha_i$     &  $Y_{7r}(0)$    & $Y_{7i}(0)$  \\ \hline
%               &            &                 &             &          \\ 
 $1$ & $0.29324967$ & $-5.9321327\times 10^{-3}$ & $-0.1728\times 10^{-8}$ & $-0.2485\times 10^{-8}$  \\
 $2$ & $0.17673123$ & $0.12151321$              & $-0.5690\times 10^{-5}$ &  $-0.1086\times 10^{-4}$  \\      
 $3$ & $0.21522309$ & $0.21439414$              & $-0.2256\times 10^{-4}$ & $0.2402\times 10^{-4}$    \\
 \hline
\end{tabular} 
\end{center}
\caption{The spatial eigenvalues obtained after Newton-Raphson polishing tabulated for $Re=1500$ and $\omega_r=0.1$ for 2D zero-pressure gradient incompressible and compressible boundary layer at $M=0.1$ over an adiabatic wall. The corresponding values of the real and imaginary part of the dispersion relations, after achieving convergence, are also enlisted.}
\label{table_1}
\end{table}

We first compare the spatial stability analysis of incompressible $2D$ zero-pressure gradient (ZPG) Blasius boundary layer and compressible boundary layer on the insulated wall at $M=0.1$ for the validation of the developed approach using CMM. We choose a Reynolds number of $Re=1500$ and $\omega_r=0.1$. The stability analysis of the incompressible Blasius boundary layer is carried out by solving the Orr-Sommerfeld equation (OSE)\cite{drazin2004} 
\begin{eqnarray}
  \varphi''''-2\Delta^2 \varphi''+\Delta^4\varphi = iRe\left[Q\left(\varphi''-\Delta^2 \varphi\right)-\alpha \frac{d^2\bar{U}}{dy^2} \varphi \right]  \label{ose} 
\end{eqnarray}

\noindent Equation~(\ref{ose}) shows that OSE is a forth-order ODE. When CMM is applied to the OSE, the corresponding dispersion relation is obtained as the no-slip and the zero-normal boundary condition at the wall. These conditions lead to the dispersion relation given as $Y_1=\left(\varphi_1\varphi_3'-\varphi_1'\varphi_3\right)=0$ at the wall, where $\varphi_1$ and $\varphi_3$ are the independent modes of the OSE, which decays in the free-stream\cite{drazin2004,sengupta2012}. In Fig.~\ref{fig4}(a), we plot the contours of $Y_{1r}(0)=0$ and $Y_{1i}(0)=0$ in the $(\alpha_r,\alpha_i)$-plane as obtained by integrating the CMM equation corresponding to OSE from free-stream to the wall. The intersection of the contour lines corresponding to $Y_{1r}(0)=0$ an $Y_{1i}(0)=0$ indicates a spatial eigenvalue for OSE. We note the existence of three intersection points, which are noted as modes-$1$, $2$, and $3$, respectively. The obtained values, noted in Fig.~\ref{fig4}(a) for the three modes, can be further polished by the Newton-Raphson method using CMM. These values are noted in Table~\ref{table_1}(a). The value of $\alpha_i$ for mode-$1$ is noted to be negative, indicating this mode displays spatial instability, and disturbances corresponding to this mode grow in space as $e^{-\alpha_i x}$. The other two modes are spatially stable.

Figure~\ref{fig4}(b) shows the contours lines of $Y_{7r}(0)=0$ and $Y_{7i}(0)=0$ when CMM is applied to a compressible boundary layer on the adiabatic plate for $M=0.1$. The dispersion relation is given as $Y_{7}(0)=0$ as discussed in Sec.~\ref{sec_IIB} and therefore, the intersection points of $Y_{7r}(0)=0$ and $Y_{7i}(0)=0$ indicate eigenmodes. Figure~\ref{fig4}(b) shows that modes-$1$, $2$, and $3$ are clearly noted. These values are subsequently polished using the Newton-Raphson method as tabulated in Table~\ref{table_1}(b). The convergence is achieved when $|\delta\alpha|=|\alpha_{n+1}-\alpha_n|<10^{-16}$ and $|\delta Y_7(0)|=|Y_{7,n+1}(0)-Y_{7,n}(0)|<10^{-16}$, where the subscript $n$ indicates the $n^{th}$ iteration of the Newton-Raphson method. Figure~\ref{fig4}(b) also shows that there are significant numerical disturbances when $\alpha_r<0.14$. Even if several crossings are noted on the top part of the domain, these are spurious as no convergence in the Newton-Raphson method corresponding to these values is achieved. We note that modes-$1$ and $2$ are similar for incompressible and compressible boundary layers. Mode-$3$, for the compressible boundary layer, has a higher value of $\alpha_r$ than the incompressible one. Like the incompressible one, the mode-$1$ for the compressible boundary layer displays spatial instability, and modes-$2$ and $3$ are spatially stable.

We find that for both 2D incompressible and $M=0.1$ compressible boundary layer on the insulated wall, modes-$2$ and $3$ are always spatially stable when $\omega_r$ and $Re$ is varied over a wide range, \textit{i.e.}, $\alpha_i>0$ always for these modes. These modes are also noted to disappear when $\omega_r$ and $Re$ are below certain corresponding values, as also reported in Sengupta, Nair, \& Rana \cite{sengupta1997} for ZPG incompressible boundary layer. Next, we track mode-$1$ in the $(Re,\omega_r)$-plane by performing Newton-Raphson polishing while using a neighboring polished mode as the initial guess. In Fig.~\ref{fig4}(c), we compare $\alpha_i=0$ contour for mode-$1$ for both cases. This curve is called the \textit{Neutral curve} \cite{drazin2004}. Inside this curve, $\alpha_i<0$, and the flow is spatially unstable, while outside this curve, $\alpha_i>0$, and the flow is spatially stable. We define the critical Reynolds number $Re_{cr}$ as the lowermost limiting Reynolds number below which the mode-$1$ is always stable. We note identical $Re_{cr}$ ($Re_{cr}\simeq 520$) and lower branch of the neutral curve for both 2D cases. The value of $Re_{cr}\simeq 520$ for incompressible boundary layers is reported in various literature \cite{drazin2004,tollmien1936,schmid2002,sengupta2012}. The upper branch of the neutral curve shows a slight difference due to the effect of flow compressibility. We note that the $M=0.1$ compressible boundary layer is slightly more stable than the incompressible one. In Fig.~\ref{fig4}(c,d), we further compare the group velocity $V_g$ of disturbances corresponding to mode-$1$. The group velocity is computed as $V_g=\partial \omega_r/\partial \alpha_r$ at fixed $Re$, which indicates the speed at which the wave energy travels. Figure~\ref{fig4}(c,d) shows that the group velocity contours for compressible and incompressible shear layers are almost identical.

\subsection{Effects of flow compressibility on spatial stability}\label{sec_comp}

\begin{figure}[!htbp]
    \centering
    \includegraphics[width=1.0\textwidth]{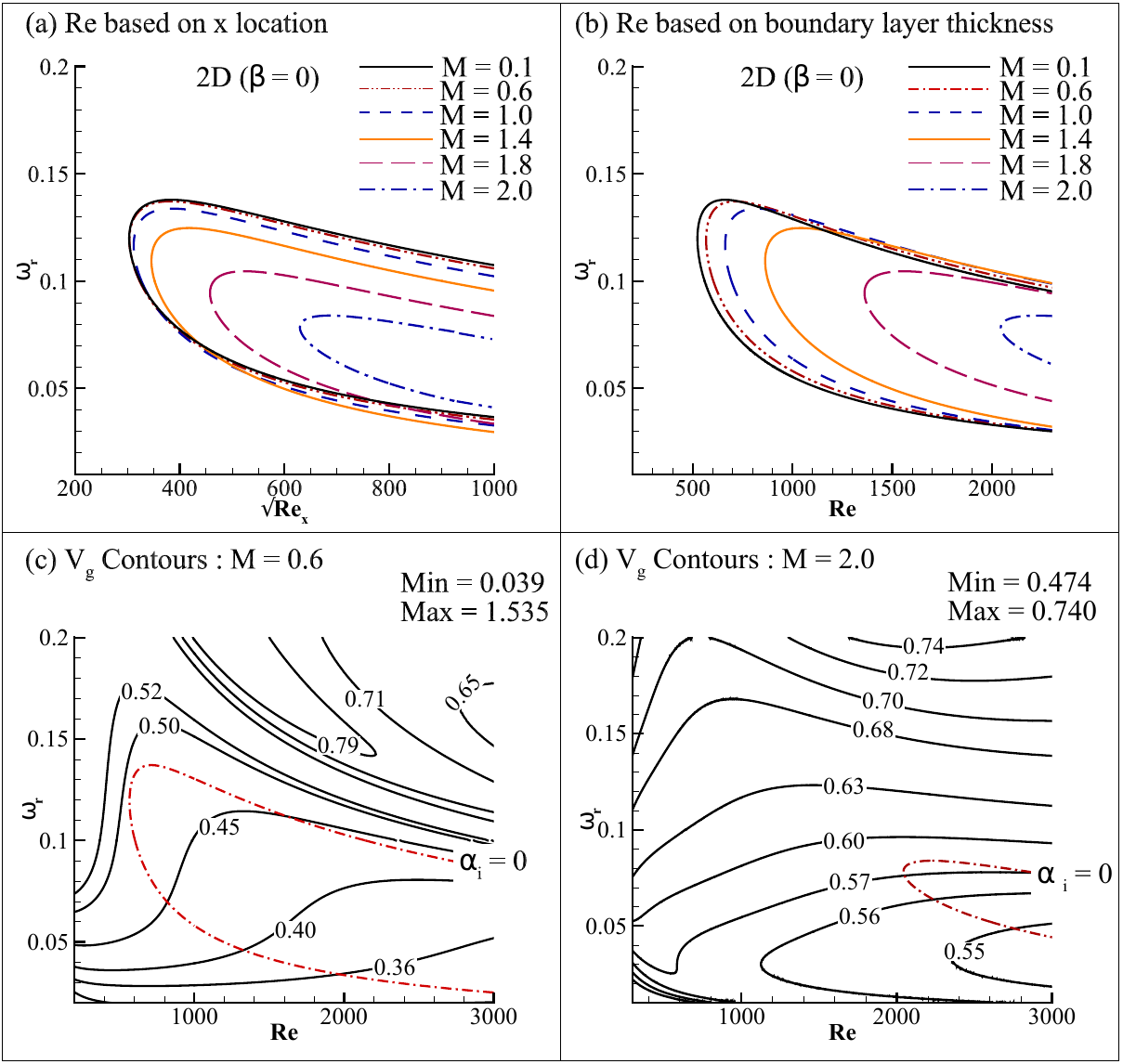}
    \caption{The Neutral curve plotted for indicated Mach number cases in (a) $(\sqrt{Re_x},\omega_r)$- and (b) $(Re,\omega_r)$-plane for $2D$ disturbance. Group velocity contours shown for (c) $M=0.6$ and (d) $M=2$ in the $(Re,\omega_r)$-plane. The corresponding neutral curve in frames (c,d) is represented as a dashed line.}
    \label{fig5}
\end{figure}

We further explore the effects of flow compressibility on the spatial stability of the compressible wall-bounded shear layer on an adiabatic wall. Here, we consider the Mach number cases ranging from $0.1$ to $2$ and only their corresponding $2D$ stability characteristics. As noted in Fig.~\ref{fig4}(a), multiple spatial eigenmodes exist for each case. However, only one mode displays spatial instability, and all other modes are spatially stable, \textit{i.e.}, $\alpha_i>0$. In Figs.~\ref{fig5}(a,b), we show the neutral curve for the only one spatially unstable mode plotted in (a) $(\sqrt{Re_x},\omega_r)$- and (b) $(Re,\omega_r)$-planes, where $Re_x$ is the Reynolds number based on local streamwise coordinate and $Re=c_{\delta} \sqrt{Re_x}$ as described in Sec.~\ref{sec_similarity}. Here, we provide two different measures as the value of $c_{\delta}$ depends on the $M$ (see Fig.~\ref{fig3}(c)), indicating $Re$ not to be a good measure of the extent of the unstable zone when different Mach number cases are compared. Figures~\ref{fig5}(a,b) show that with an increase in Mach number, the extent of the spatially unstable zone reduces. This reduction is marginal for subsonic cases (especially when viewed at the $(\sqrt{Re_x},\omega_r)$-plane) and subsequently becomes more severe as the flow becomes supersonic. An explanation in this regard was given in Mack\cite{mack1984}, which concluded that the viscous instability weakens with an increase in flow Mach number, while the dominance of inviscid instability extends to lower Reynolds numbers. We show the group velocity contours for the spatially unstable modes in Figs.~\ref{fig5}(c,d) corresponding to $M=0.6$ and $2$, respectively, which shows that at higher Mach numbers, the disturbances travel relatively faster. The contour lines become increasingly parallel while the deviation in the values becomes significantly lesser and is closer to $0.5$, \textit{i.e.}, the speed of a pure acoustic disturbance. For $M=2$, these indicate the lessening of viscous dispersive effects, and the disturbances increasingly travel following inviscid acoustic modes as postulated in Mack\cite{mack1984}.

\subsection{Effects of spanwise wavenumber $\beta$ on spatial stability} \label{sec_beta_effect}

\begin{figure}[!htbp]
    \centering
    \includegraphics[width=1.0\textwidth]{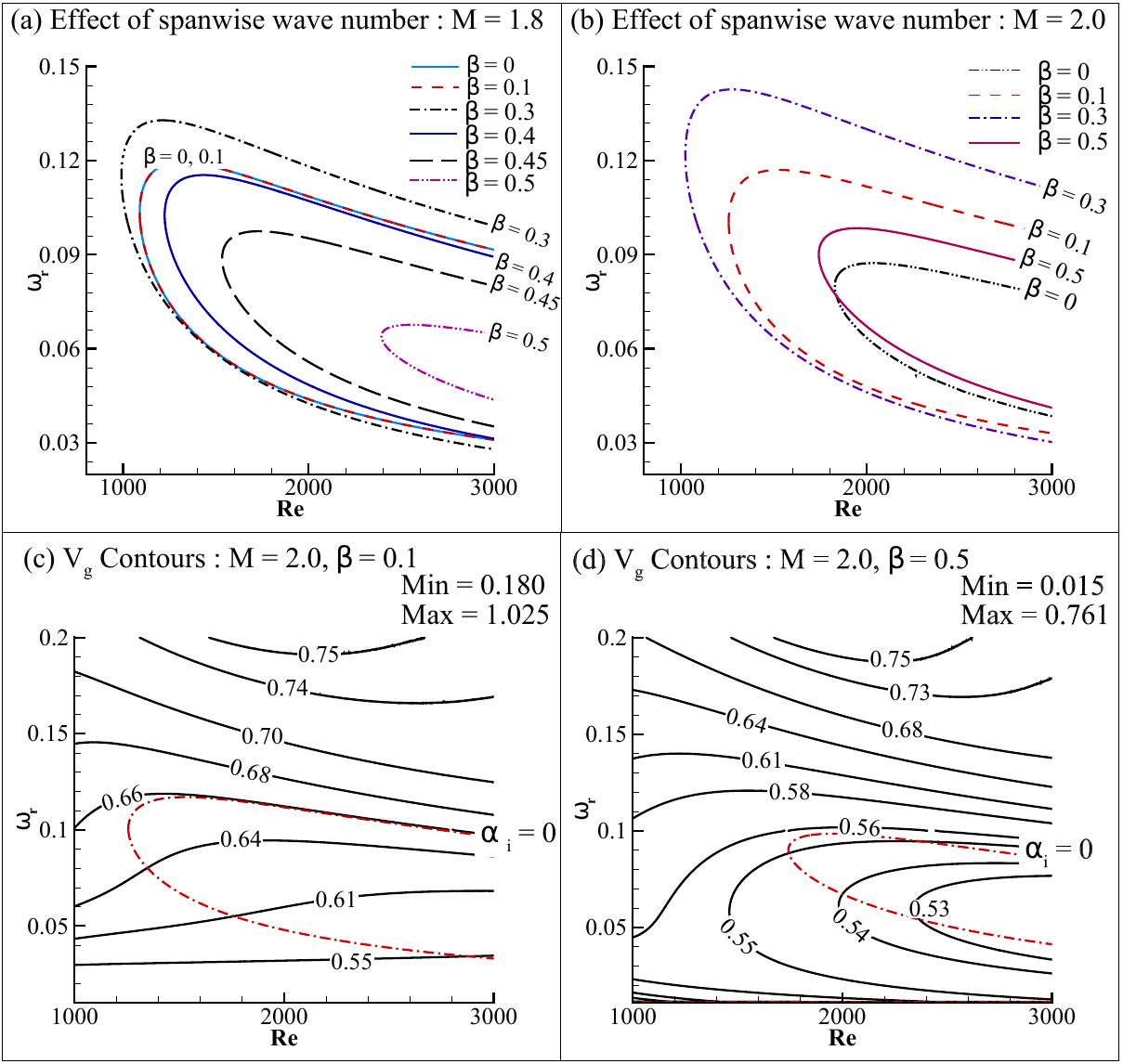}
    \caption{The Neutral curve plotted in $(Re,\omega_r)$-plane corresponding to $2D$ compressible boundary layer on insulated plate corresponding to indicated spanwise wavenumber cases for (a) $M=1.8$ and (b) $M=2$. Group velocity contours shown for $M=2$ case in $(Re,\omega_r)$-plane when (c) $\beta=0.1$ and (d) $\beta=0.5$. The corresponding neutral curve in frames (c,d) is represented as a dashed line.}
    \label{fig6}
\end{figure}

Next, we investigate the effects of spanwise wavenumber $\beta$ on the spatial stability of the compressible boundary layer over an adiabatic wall. We consider $M=1.8$ and $2$ with the spanwise wavenumber $\beta$ ranging from $0$ to $0.5$. As the problem is essentially $3D$, the CMM is applied to the $8^{th}$-order system (even for the case corresponding to $\beta=0$), which results in $72$ compound matrix equations.  

Multiple spatial eigenmodes exist for each of these cases, while only one mode displays spatial instability, and the rest are always spatially stable. In Figs.~\ref{fig6}(a,b), we show the neutral curve for the sole spatially unstable mode in $(Re,\omega_r)$-plane for $M=1.8$ and $2$, respectively. We observe that the value of the critical Reynolds number $Re_{cr}$ and the extent of the spatially unstable zone depends on $\beta$. For $M=1.8$, the neutral curve for $\beta=0$ and $\beta=0.1$ is almost indistinguishable. The subsequent increase in $\beta$ decreases $Re_{cr}$ and increases the extent of the unstable zone as noted for $\beta=0.3$ in Fig.~\ref{fig6}(a). However, a subsequent increase in $\beta$ increases $Re_{cr}$ and decreases the extent of the spatially unstable zone. Beyond $\beta=0.4$, the decrease in the extent of the unstable zone is more drastic, as noted by comparing the neutral curves for $\beta=0.4$, $0.45$, and $0.5$ in Fig.~\ref{fig6}(a). Therefore, an optimum value of $\beta$ close to $0.3$ exists for which $Re_{cr}$ is minimum, and the extent of the spatially unstable zone is maximum for $M=1.8$. Similar conclusions are also noted for $M=2$ in Fig.~\ref{fig6}(b). For $M=2$, the decrease in $Re_{cr}$ and increase in the extent in the unstable zone from $\beta=0$ to $0.3$ is more severe than $M=1.8$. Subsequently, we note an increase in $Re_{cr}$ and a decrease in the extent of the unstable zone, which is less severe than the $M=1.8$ case, as the extent of the unstable zone for $\beta=0.5$ is still larger than $\beta=0$. The group velocity contours for $\beta=0.1$ and $0.5$ corresponding to $M=2$ are depicted in Fig.~\ref{fig6}(c,d), which shows that disturbances are less dispersive for $\beta=0.1$ than $0.5$ for lower frequency components ($\omega_r \lesssim 0.15$). These components, collectively as a group, are also noted to travel slower for $\beta=0.5$ than $\beta=0.1$. 

\begin{figure}[!htbp]
    \centering
    \includegraphics[width=1.0\textwidth]{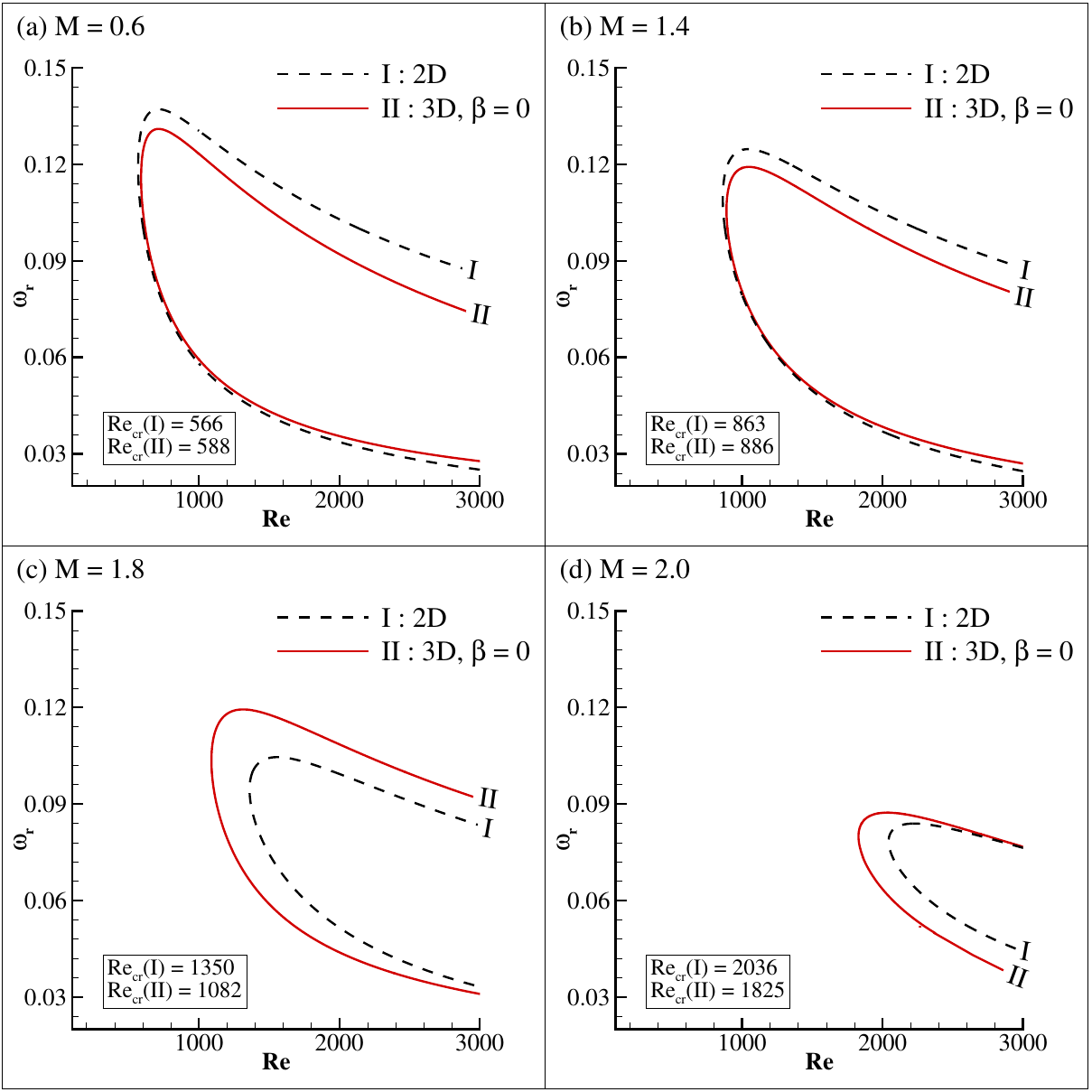}
    \caption{The neutral curve plotted in $(Re,\omega_r)$-plane corresponding to $2D$ ($6^{th}$-order system) and $3D$ disturbance case with $\beta=0$ ($8^{th}$-order system) for (a) $M=0.6$,
      (b) $M=1.4$, (c) $M=1.8$ and (d) $M=2$. The corresponding $Re_{cr}$ is also indicated in the frames.}
    \label{fig7}
\end{figure}

Regarding the stability calculation of the compressible boundary layers, there is ambiguity regarding the appropriate order of the equation. For the incompressible flows, whether the disturbances are $2D$ or $3D$ ($\beta\ne 0$ or $\hat{w} \ne 0$), the stability equations are given by the appropriate form of the $4^{th}$-order OSE given by Eq.~(\ref{ose}). For the compressible boundary layers, such simplifications are not straight-forward, as also noted in Mack\cite{mack1984}, Dunn \& Lin \cite{dunn1955} and Lees \& Reshotko \cite{lees1962}. Even when $\beta=0$, the $3D$ compressible stability equations are not identical to the $2D$ version. Considering Eqs.~(\ref{app1_eq1}-\ref{app1_eq5}) we note that $\Omega$ contributes only to the equation for $\Theta$ through the dissipation term $\mathcal{T}_1=\left[2 \beta \frac{\bar{\mu}}{Re}\left(\gamma-1\right)\frac{M^2}{\Delta^2} \frac{d\bar{U}}{dy} \right]$. Mack\cite{mack1984} proposed to simply set this term to zero while using the resultant $6^{th}$-order system for eigenvalue calculations so that the accuracy is improved specifically at higher supersonic Mach numbers. We note from Eqs.~(\ref{app1_eq1}-\ref{app1_eq5}) that for the $2D$ case, we only need to satisfy Eqs.~(\ref{app1_eq1}-\ref{app1_eq4}) with $\mathcal{T}_1=0$ corresponding to $\psi(0)=\varphi(0)=\Theta^{\prime}(0)=0$, while for the $3D$ case with $\beta=0$, we also need to satisfy an additional decoupled equation  

\begin{eqnarray}
  \left[ \frac{\bar{\mu}}{Re} \right] \Omega^{\prime\prime}  
  + \left[ \frac{C_1}{Re}\right] \Omega^{\prime} +
  \left[ - \frac{\bar{\mu}}{Re} \Delta^2 - i\bar{\varrho}Q \right] \Omega = 0 
  \label{app1_eq5a}
\end{eqnarray}

\noindent along with $\Omega(0)=0$. Therefore, it can be deduced that an eigenvalue for the $2D$ case will represent the eigenvalue for the $3D$ case with $\beta=0$ corresponding to the degenerative trivial condition of $\hat{w}=0$ or $\Omega=0$. An eigenvalue for the $3D$ case with $\beta=0$ should not necessarily represent an eigenvalue for the respective $2D$ case, as for the former, non-trivial Squire mode is not imposed at free-stream while using CMM. These points are illustrated next using Fig.~\ref{fig7} and Table~\ref{table_2}.

%%%% Insert Table-2

\begin{table}[h]
    \centering
    \begin{tabular}{|c|c|c|c|c|c|}\hline
        \multicolumn{4}{c}\text{(a) $Re=2500$, $\omega_r =0.06$} \\ \hline
        $M$  & Formulation & $\alpha_r$ & $\alpha_i$ & $Y_{jr}(0)$  & $Y_{ji}(0)$ \\ \hline
%             &             &            &            &          &         \\ 
        \multirow{2}{*}{$0.6$}
        & $2D$                & $0.18893069$ & $-9.41920470 \times 10^{-3}$ & $3.6520 \times 10^{-9}$ & $1.5799 \times 10^{-9}$ \\
        & $3D$ with $\beta=0$ & $0.18952319$ & $-6.22944070 \times 10^{-3}$ & $1.2104 \times 10^{-9}$ & $1.4429 \times 10^{-9}$ \\
        \hline
        \multirow{2}{*}{$1.4$}
        & $2D$                & $0.14093782$ & $-3.36921910 \times 10^{-3}$ & $1.4945 \times 10^{-8}$ & $-1.5668 \times 10^{-8}$ \\
        & $3D$ with $\beta=0$ & $0.14325045$ & $-2.94359420 \times 10^{-3}$ & $1.5272 \times 10^{-10}$ & $-3.8199 \times 10^{-10}$ \\
        \hline
        \multirow{2}{*}{$1.8$}
        & $2D$                & $0.11736789$ & $-6.51037790 \times 10^{-4}$ & $-2.9220 \times 10^{-9}$ & $-6.0412 \times 10^{-10}$ \\
        & $3D$ with $\beta=0$ & $0.12099984$ & $-8.21889200 \times 10^{-4}$ & $1.4344 \times 10^{-9}$  & $1.8298 \times 10^{-9}$ \\
        \hline
        \multirow{2}{*}{$2.0$}
        & $2D$                & $0.10879819$ & $-7.24156630 \times 10^{-5}$ & $-1.4542 \times 10^{-9}$ & $-2.3321 \times 10^{-9}$ \\
        & $3D$ with $\beta=0$ & $0.11232695$ & $-2.36114590 \times 10^{-4}$ & $1.2846 \times 10^{-9}$  & $-1.4493 \times 10^{-10}$ \\
        \hline
 \multicolumn{4}{c}\text{(b) $Re=2500$, $\omega_r =0.1$}  \\ \hline
        $M$  & Formulation & $\alpha_r$ & $\alpha_i$ & $Y_{jr}(0)$ & $Y_{ji}(0)$ \\ \hline
 %            &             &            &            &          &        \\ 
        \multirow{2}{*}{$0.6$}
        & $2D$                & $0.28802121$ & $4.43784240 \times 10^{-3}$ & $2.1877 \times 10^{-9}$ & $2.7341 \times 10^{-8}$ \\
        & $3D$ with $\beta=0$ & $0.29012486$ & $1.39401840 \times 10^{-2}$ & $-4.9300 \times 10^{-10}$ & $-9.2410 \times 10^{-10}$ \\
        \hline
        \multirow{2}{*}{$1.4$}
        & $2D$                & $0.22044748$ & $1.0555690 \times 10^{-3}$ & $-2.4435 \times 10^{-9}$ & $-4.6050 \times 10^{-9}$ \\
        & $3D$ with $\beta=0$ & $0.22088128$ & $3.2584926 \times 10^{-3}$ & $1.1763 \times 10^{-9}$ & $-9.2065 \times 10^{-10}$ \\
        \hline
        \multirow{2}{*}{$1.8$}
        & $2D$                & $0.18988059$ & $7.01090500 \times 10^{-4}$ & $-2.3585 \times 10^{-8}$ & $-9.6791 \times 10^{-9}$ \\
        & $3D$ with $\beta=0$ & $0.19109581$ & $1.28319800 \times 10^{-3}$ & $3.6364 \times 10^{-11}$ & $4.5576 \times 10^{-10}$ \\
        \hline
        \multirow{2}{*}{$2.0$}
        & $2D$                & $0.17793112$ & $6.38880180 \times 10^{-4}$ & $7.0879 \times 10^{-9}$ & $-8.7970 \times 10^{-9}$ \\
        & $3D$ with $\beta=0$ & $0.17952602$ & $8.81986290 \times 10^{-4}$ & $1.5717 \times 10^{-9}$ & $-1.1064 \times 10^{-9}$ \\
        \hline
    \end{tabular}
    \caption{The spatial eigenvalues corresponding to the $2D$ and $3D$ case with $\beta=0$ tabulated for $M=0.6$, $1.4$, $1.8$ and $2$ when $Re=2500$ and (a) $\omega_r=0.06$ and
      (b) $\omega_r=0.1$. The corresponding values of the real and imaginary part of the wall dispersion function $Y_j(0)$ are also enlisted, where $j=7$ and $23$ for $2D$ and $3D$ cases,
      respectively.}
    \label{table_2}
\end{table}

%%%% Insert Table-2a

\begin{table}[htbp]
    \centering
    \renewcommand{\arraystretch}{1.5}
    \begin{tabular}{|c|c|c|c|c|c|c|}
        \hline
        \multirow{2}{*}{Formulation} & \multicolumn{2}{c|}{$M=0.6$} & \multicolumn{2}{c|}{$M=1.8$} & \multicolumn{2}{c|}{$M=2$} \\
        \cline{2-7}
         & $Re_{cr}$ & $\sqrt{Re_{x_{cr}}}$ & $Re_{cr}$ & $\sqrt{Re_{x_{cr}}}$ & $Re_{cr}$ & $\sqrt{Re_{x_{cr}}}$ \\
        \hline
          $2D$               & $566.35$  & $303.51$ & $1360.70$ & $458.15$  & $2043.58$ & $630.73$ \\
        $3D$ with $\beta=0$  & $588.12$  & $315.17$ & $1089.19$ & $366.73$  & $1829.06$ & $564.52$ \\
        $\beta=0.1$          & $617.32$  & $330.82$ & $1089.05$ & $366.68$  & $1256.57$ & $387.83$ \\
        $\beta=0.3$          & $1214.11$ & $650.65$ & $990.23$  & $333.41$  & $1022.91$ & $315.71$ \\
        $\beta=0.4$          &    $-$    & $-$      & $1223.46$ & $411.93$  & $1144.14$ & $353.12$ \\
        $\beta=0.45$         &    $-$    & $-$      & $1533.75$ & $516.41$  & $1329.31$ & $410.28$ \\
        $\beta=0.5$          &    $-$    & $-$      & $2392.37$ & $805.51$  & $1742.46$ & $537.79$ \\
        \hline
    \end{tabular}
            \caption{$Re_{cr}$ based on displacement thickness and local streamwise coordinate is tabulated for $2D$ compressible boundary layer corresponding to indicated disturbance cases.}
    \label{table_2a}
\end{table}

In Fig.~\ref{fig7}(a-d), we show the neutral curve in $(Re,\omega_r)$-plane corresponding to $2D$ ($6^{th}$-order system) and $3D$ case with $\beta=0$ ($8^{th}$-order system) for $M=0.6$, $1.4$, $1.8$, and $2$, respectively. In Table~\ref{table_2}, we show the values of the eigenvalue $\alpha$ for $Re=2500$ and $\omega_r=0.06$ and $0.1$, respectively, for the above cases. The first point, \textit{i.e.}, $Re=2500$ and $\omega_r=0.06$ is inside the neutral curve, while the second point, \textit{i.e.}, $Re=2500$ and $\omega_r=0.1$ is in the spatially stable region for all the cases. The converged value of the dispersion function $Y_j(0)$ is also noted in the table, where $j=7$ and $23$ for $2D$ and $3D$ cases, respectively. Figure~\ref{fig7} shows that for the subsonic to low supersonic Mach numbers (up to at least $M=1.4$), the unstable zone corresponding to the $2D$ case is larger along with a slightly lower value of $Re_{cr}$. A tentative reason may be provided following Mack\cite{mack1984} that the viscous modes dominate for these cases, and as non-zero perturbation spanwise velocity $\hat{w}$ needs to extract energy from the mean flow for sustenance, the stability of the flow is enhanced. However, for $M=1.8$ and $2$ cases, the $3D$ case with $\beta=0$ is more spatially unstable than the corresponding $2D$ case. This is also noted for the corresponding $\alpha$'s tabulated in table~\ref{table_2} for $Re=2500$ and $\omega_r=0.06$. A possible reason for such behavior may be the dominance of apparently inviscid mode for higher Mach numbers as pointed in Mack\cite{mack1984}, and Lees \& Reshotko \cite{lees1962}. These points need further detailed investigation, possibly by comparing the wall-normal eigenstructure.    

We have also investigated the effect of $\beta$ for a subsonic Mach number of $M=0.6$ (results not presented here). For the subsonic case, stability of the flow increases with an increase in spanwise wavenumber both in terms of reduction in $Re_{cr}$ and the extent of the spatially unstable zone in $(Re,\omega_r)$-plane. In Table~\ref{table_2a}, we tabulate the critical Reynolds number $Re_{cr}$ based on displacement thickness and local streamwise coordinate for $M=0.6,1.8$ and $2$. It is often concluded that a $2D$ mean flow is more stable for $3D$ disturbances than pure $2D$ disturbances (see \textit{i.e.}, Drazin \& Reid \cite{drazin2004}). While this conclusion is valid for incompressible boundary layers\cite{maddipati2021} and also for subsonic compressible adiabatic boundary layer cases, as noted here for $M=0.6$, Figs.~\ref{fig6} and \ref{fig7} and Tables~\ref{table_2} and \ref{table_2a} prove that, it strictly does not hold for supersonic boundary layer cases.  

% For $M=1.8$ and $2$, we note $\left(\alpha_i\right)_{\text{$3D$ with $\beta=0$}}<\left(\alpha_i\right)_{2D}<0$. 

\subsection{Spatial stability corresponding to $2D$ disturbances for $M>3$} \label{sec_M_ge_3}

Next, we investigate the spatial instability of compressible boundary layers for $M>3$. Mack\cite{mack1984,mack1984pof} reported the appearance of the second unstable mode at finite Reynolds number for $M>3$, which is described to be inviscid in nature due to the presence of generalized inflection point $y_s=(U'/T')'$ at certain height. The neutral curve for the second mode is at the top of that corresponding to the first viscous mode in Mack \cite{mack1984} for $3<M<4.5$ insulated wall-bounded boundary layer. As $M$ increases, the amplification rate of the second mode increases while the corresponding neutral curve tends to merge with that for the viscous first mode. The merger between the first and the second mode happens according to Mack\cite{mack1984} beyond $M=4.6$ for the adiabatic wall-bounded boundary layer. The existence of the high-frequency second mode is also reported in Kendall\cite{kendall1975}, Stetson\cite{stetson1992}, Fedorov \& Tumin \cite{fedorov2003} and other references contained therein. The above observations motivate us to relook at the $2D$ spatial instability of the supersonic boundary layers beyond $M>3$ by CMM.

\begin{figure}[!htbp]
    \centering
    \includegraphics[width=0.95\textwidth]{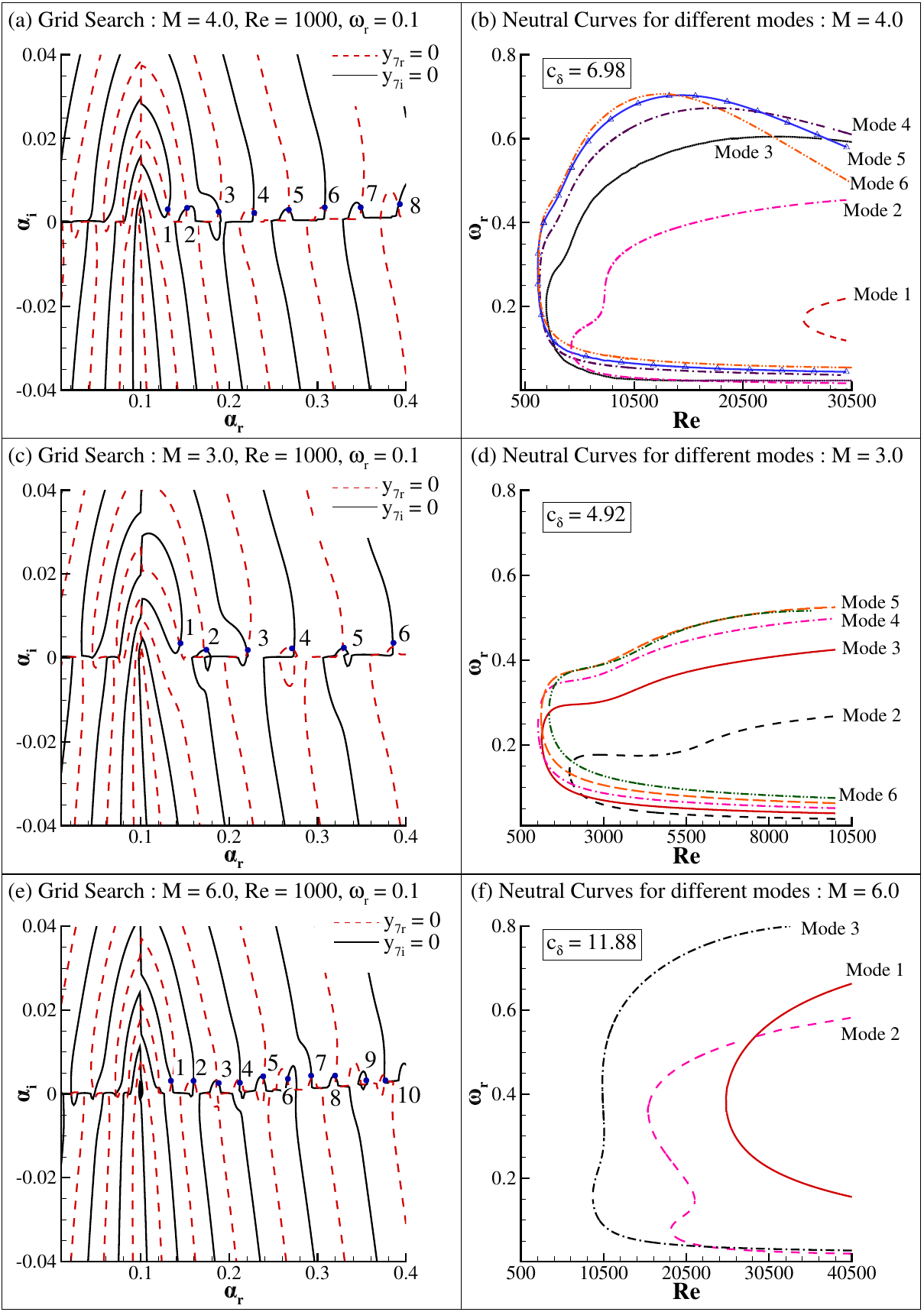}
    \vspace{-0.5cm}
    \caption{Spatial $2D$ modes in $(\alpha_r,\alpha_i)$-plane for $Re=1000$ and $\omega_r=0.1$ and selected neutral curves shown for (a,b) $M=4$, (c,d) $M=3$ and (e,f) $M=6$ for $2D$ compressible boundary layer over an insulated plate.}
    \label{fig8}
\end{figure}

\begin{table}
	\vspace{-0cm}
	\begin{center}
		\begin{tabular}{| c | c | c | c | c | c |} % {| c| c| c| c| c| c| c| c| c |}
			\hline 
			% \multicolumn{5}{c} \text{(a) $Re=1000$, and $\omega_r =0.1$}  \\ \hline              
			& Mode No. & $\alpha_r$ & $\alpha_i$       & $Y_{7r}(0)$          & $Y_{7i}(0)$ \\ \hline
			%   &            &                  &                  &          \\ \hline 
			\multirow{8}{*}{\text{(a) $Re=1000$,}} 
			& $1$  & $0.13099846$ & $2.61582640\times 10^{-3}$ & $-6.4153\times 10^{-9}$  & $-6.9270\times 10^{-9}$   \\
			\multirow{8}{*}{\text{and $\omega_r =0.1$}}
			& $2$  & $0.15160470$ & $2.94952140\times 10^{-3}$ & $-9.7098\times 10^{-9}$  & $-2.2834\times 10^{-9}$  \\
			& $3$  & $0.18859601$ & $2.67518660\times 10^{-3}$ & $-2.1216\times 10^{-9}$  & $-1.9402\times 10^{-8}$  \\
			& $4$  & $0.22759297$ & $2.65627840\times 10^{-3}$ & $8.5090\times 10^{-9}$  & $2.6440\times 10^{-9}$   \\
			& $5$  & $0.26758295$ & $2.79169690\times 10^{-3}$ & $-1.2911\times 10^{-8}$ & $9.8854\times 10^{-10}$  \\
			& $6$  & $0.30832326$ & $3.03706200\times 10^{-3}$ & $-1.1347\times 10^{-8}$  & $2.6369\times 10^{-8}$  \\
			& $7$  & $0.34962437$ & $3.35736340\times 10^{-3}$ & $-2.7888\times 10^{-8}$  & $-3.6595\times 10^{-9}$  \\
			& $8$  & $0.39132136$ & $3.73667760\times 10^{-3}$ & $-3.7469\times 10^{-8}$ & $2.0684\times 10^{-9}$  \\
			\hline
			% \multicolumn{5}{c}  \text{} \\
			% \multicolumn{5}{c} \text{(b) $Re=15000$, and $\omega_r =0.15$} \\ \hline
			%      & Mode No.     & $\alpha_r$  & $\alpha_i$     &  $Y_{7r}$    & $Y_{7i}$  \\ \hline
			%            &            &                 &             &          \\ \hline
			\multirow{8}{*}{\text{(b) $Re=15000$,}} 
			& $1$  & $0.19498196$ & $1.93222380\times 10^{-4}$  & $-1.0033\times 10^{-8}$  & $7.4700\times 10^{-9}$   \\
			\multirow{8}{*}{\text{and $\omega_r =0.15$}}
			& $2$  & $0.21168292$ & $-2.76303440\times 10^{-4}$ & $5.4583\times 10^{-9}$  & $-3.6451\times 10^{-9}$  \\
			& $3$  & $0.24550970$ & $-7.74741060\times 10^{-4}$ & $-4.8893\times 10^{-10}$ & $9.6411\times 10^{-10}$  \\
			& $4$  & $0.28347549$ & $-1.00149330\times 10^{-3}$ & $-5.3637\times 10^{-8}$  & $7.0336\times 10^{-9}$   \\
			& $5$  & $0.32309651$ & $-1.08311510\times 10^{-3}$ & $1.1205\times 10^{-9}$ & $7.0153\times 10^{-9}$  \\
			& $6$  & $0.36358333$ & $-1.06486810\times 10^{-3}$ & $-5.1550\times 10^{-9}$  & $-4.2671\times 10^{-9}$  \\
			& $7$  & $0.40459979$ & $-9.47628640\times 10^{-4}$ & $-5.1089\times 10^{-9}$  & $-1.1243\times 10^{-8}$  \\
			& $8$  & $0.44597900$ & $-7.45557130\times 10^{-4}$  & $-2.8856\times 10^{-8}$ & $2.5628\times 10^{-8}$  \\
			\hline
			% \multicolumn{5}{c}  \text{} \\
			% \multicolumn{5}{c} \text{(c) $Re=29000$, and $\omega_r =0.15$} \\ \hline
			%      & Mode No.  & $\alpha_r$  & $\alpha_i$     &  $Y_{7r}$    & $Y_{7i}$  \\ \hline 
			%             &            &                 &             &          \\ \hline
			\multirow{8}{*}{\text{(c) $Re=29000$,}} 
			& $1$  & $0.19497371$ & $-2.6368667\times 10^{-5}$ & $-2.0988\times 10^{-9}$  & $-4.6073\times 10^{-10}$   \\
			\multirow{8}{*}{\text{and $\omega_r =0.15$}}
			& $2$  & $0.21163477$ & $-3.9839061\times 10^{-4}$ & $-1.6398\times 10^{-8}$  & $-2.3914\times 10^{-8}$  \\
			& $3$  & $0.24537614$ & $-8.3571899\times 10^{-4}$ & $-6.9107\times 10^{-9}$   & $2.6590\times 10^{-9}$  \\
			& $4$  & $0.28324479$ & $-9.1820088\times 10^{-4}$ & $1.2776\times 10^{-8}$  & $1.6192\times 10^{-8}$   \\
			& $5$  & $0.32281691$ & $-7.9247856\times 10^{-4}$ & $-1.9835\times 10^{-9}$  & $-2.8773\times 10^{-9}$  \\
			& $6$  & $0.36329904$ & $-5.8810069\times 10^{-4}$ & $6.9281\times 10^{-10}$   & $-3.2294\times 10^{-9}$  \\
			& $7$  & $0.40427428$ & $-3.4727529\times 10^{-4}$ & $9.0748\times 10^{-9}$  & $2.0358\times 10^{-9}$  \\
			& $8$  & $0.44549695$ & $-5.2784686\times 10^{-5}$ & $-1.6727\times 10^{-8}$  & $-2.0506\times 10^{-8}$  \\
			\hline
		\end{tabular} 
	\end{center}
	\vspace{-0.5cm} 
	\caption{The spatial eigenvalues corresponding to the indicated modes for $M=4$ obtained after Newton-Raphson polishing tabulated for (a) $(Re,\omega_r)=(1000,0.1)$, (b) $(Re,\omega_r)=(15000,0.15)$, and (c) $(Re,\omega_r)=(29000,0.15)$ for the $2D$ compressible boundary layer over insulated wall. The corresponding values of the real and imaginary parts of $Y_7(0)$ are also enlisted after convergence is achieved.}
	\label{table_3}
\end{table}

Figure~\ref{fig8}(a) shows the location of spatial modes for $M=4$, $Re=1000$ and $\omega_r=0.1$ in $(\alpha_r,\alpha_i)$-plane. The mode-$1$ is viscous in nature, while on the right of that, we note a train of modes with increasing $\alpha_r$. Here, we number these modes as $2-8$. In the figure, the range of $\alpha_r$ is taken up to $0.4$. As we increase the range of $\alpha_r$, more and more such modes are noted to exist. Modes $1-8$ for $Re=1000$ and $\omega_r=0.1$ are also tabulated in Table~\ref{table_3}(a). Figure~\ref{fig8}(b) shows the neutral curves for modes $1-6$. The shape of the neutral curves for modes-$7$ and $8$ are similar to modes $3$ and $8$ (not shown here). The $Re_{cr}$ for mode-$1$ is considerably higher than that of other modes. While $Re_{cr}$ is noted to be approximately $26051$ and $6500$ for modes $1$ and $2$, respectively, $Re_{cr}\simeq 3500$ corresponding to modes $3-8$. From the neutral curve for mode-$2$ onwards, we also note that these modes are spatially unstable over a broader range of frequencies than mode-$1$, subsonic, and low supersonic Mach number ($M \le 2$) cases. For the last cases, disturbances corresponding to $\omega_r>0.15$ are noted to be spatially stable. From mode-$2$ onwards, the maximum limit of $\omega_r$ up to which spatial instability may occur increases from $0.4$ to approximately $0.7$ for $M=4$. We also note the upper branch of the neutral curve for modes $4$, $5$, and $6$ have a dropping tendency at high Reynolds numbers. Whether these curves drop continuously and subsequently close at certain higher $Re$ will be investigated in the future. We also enlist spatial eigenvalues corresponding to modes $1-8$ obtained after Newton-Raphson polishing in Table~\ref{table_3} for $Re=15000$ and $29000$, respectively, when $\omega_r=0.15$ for reference. The first point is inside the spatially unstable zone corresponding to modes $2-8$ but outside the neutral curve for mode-$1$. The second point is within the unstable zone corresponding to modes $1-8$. We note by comparing the values of $\alpha_i$ from Tables~\ref{table_1}-\ref{table_3} that values of $\alpha_i$'s corresponding to spatial instability for $M=4$ are roughly one order of magnitude lower than subsonic or low supersonic ($M<2$) cases.

%%%% --------------- Fig. 9 ------------------------

\begin{figure}[!htbp]
    \centering
    \includegraphics[width=0.95\textwidth]{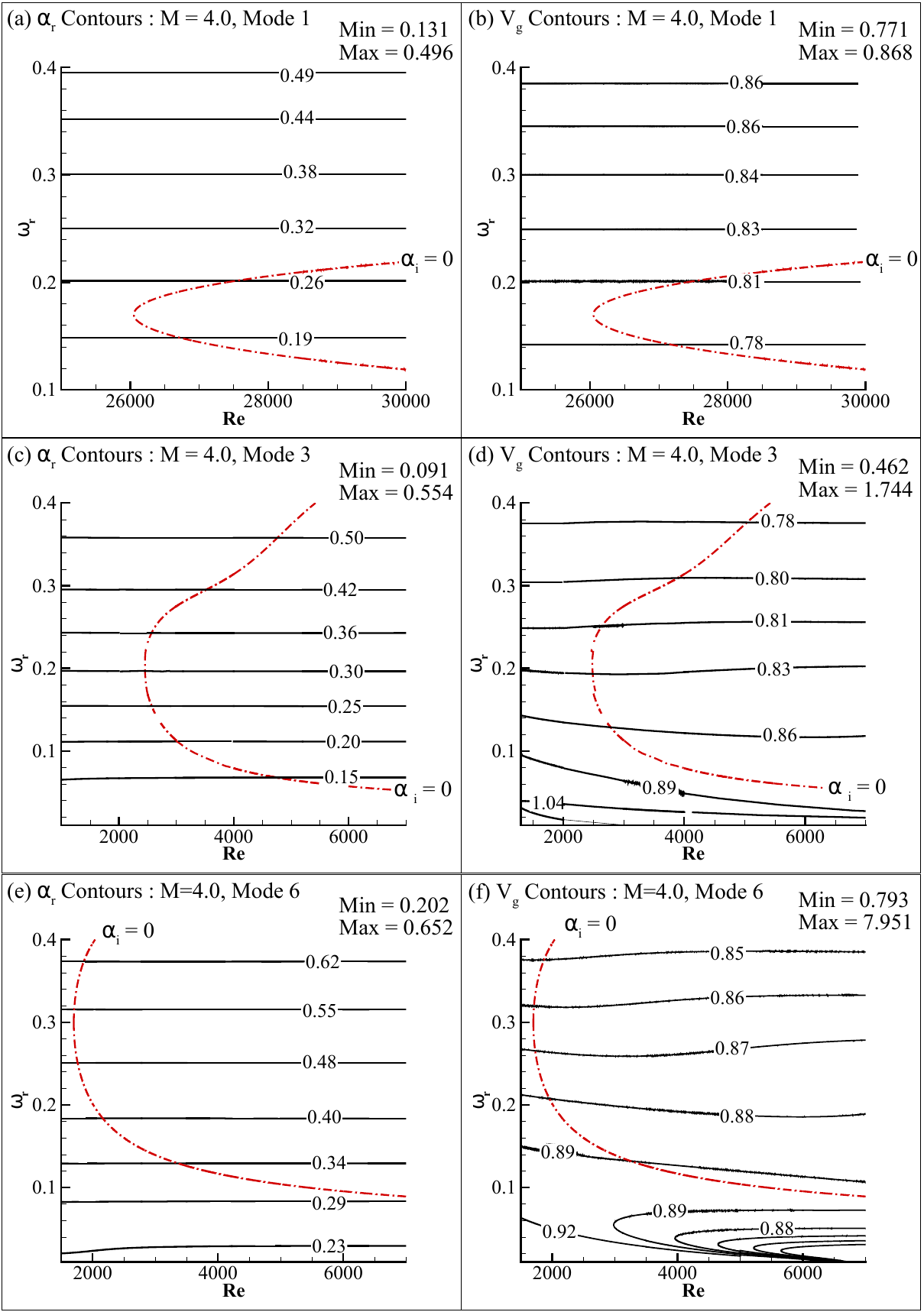}
    \vspace{-0.5cm}
    \caption{Contours of $\alpha_r$ and group velocity plotted in $(Re,\omega_r)$-plane for Modes-$1$, $3$ and $6$ corresponding to $M=4$. The corresponding neutral curve is also shown for
      reference.}
    \label{fig9}
\end{figure}

The number of such modes increases with the increase in Mach number for $M>3$. This can be concluded by comparing Figs.~\ref{fig8}(c,e) with Fig.~\ref{fig8}(a). We note only $6$ modes up to $\alpha_r=0.4$ for $M=3$, whereas $8$ and $10$ modes are noted within this range for $M=4$ and $6$, respectively. All these modes depict spatial instability beyond respective critical $Re$. The $Re_{cr}$ for mode-$1$ corresponding to $M=4$ and $6$ is $26,051$ and $25305$, respectively while that for $M=3$ is approximately $6.52\times 10^{5}$ (results not shown here). The selected neutral curves for $M=3$ and $6$ are shown in Figs.~\ref{fig8}(d) and \ref{fig8}(f), respectively, which show that the maximum limit for $\omega_r$ up to which spatially unstable disturbances are noted, increases from $M=3$ to $6$. Therefore, in summary, for $M>3$, the number of unstable modes increases drastically with the Mach number. This also coincides with the increase in the maximum limit for $\omega_r$ for which spatially unstable perturbations are observed.

It is pertinent to note that the shape of the neutral curves for mode-$2$ corresponding to $M=4$ and $6$ resembles the fused neutral curve shown for $M=4.8$ in Mack\cite{mack1984}. Except for this similarity, the nature and number of spatially unstable zones obtained here by CMM are different than that reported in Mack\cite{mack1984} and other subsequent references to date for the compressible boundary layer. To the best of our knowledge, the existence of such higher-order modes from viscous calculations for supersonic compressible boundary layers has yet to be reported in the literature so far. This is one of the novel aspects of the present investigation.

Next, we illustrate the nature of the spatial modes for higher supersonic Mach number cases. Figure~\ref{fig9} shows contours of $\alpha_r$ and group velocity $V_g=\partial \omega_r/\partial \alpha_r$ in $(Re,\omega_r)$-plane for modes-$1$, $3$ and $6$ corresponding to $M=4$. We note that $\alpha_r$ and $V_g$ are almost independent of $Re$, increasing with the increase in $\omega_r$ for mode-$1$. Almost similar variation in $\alpha_r$ is also noted for modes-$3$ and $6$; however, for these modes, a comparatively higher rate of increase in $\alpha_r$ with $\omega_r$ is observed. The group velocity $V_g$ is almost independent of $Re$ for moderate to high-frequency cases ($\omega_r\gtrsim 0.1$) corresponding to these modes similar to mode-$1$. However, in contrast to mode-$1$, $V_g$ decreases with an increase in $\omega_r$, indicating signals corresponding to these modes tend to move slower with increased frequency. The behavior displayed by modes-$1$, $3$, and $6$ is also observed for other spatially unstable modes for $M=3$, $4$, and $6$ (results not shown here). Following these variations, tentatively, we say that $\alpha_r \simeq C \omega_r^q$, where the index $q<1$ for mode-$1$ and $q>1$ for higher order modes (mode-$2$ onwards). Considering all the modes collectively, we also can say that for $M=4$, the spatially unstable disturbances travel downstream at speeds varying from $0.78$ to $0.89$ times the free-stream speed. These values are considerably higher than those compared to subsonic cases, where low to moderate frequency disturbances are seen to travel at speeds not exceeding $0.5$ times $U_{\infty}$.

\section{Summary and conclusions} \label{sec_conclusion}

We revisit the spatial linear stability analysis of the $2D$ compressible boundary layer over an adiabatic flat plate to $2D$ and $3D$ disturbances. We apply the compound matrix method (CMM) in this regard to remove the stiffness of the problem contrary to conventional methods using Gram-Schmidt ortho-normalization or discretizing the governing equations by appropriate finite difference schemes. The CMM has not been employed before to perform the stability analysis of compressible subsonic or supersonic boundary layers for $2D$ or $3D$ perturbations. While the former yields a $6^{th}$-order system, the latter results in a $8^{th}$-order system governing the wall-normal variation of spectral amplitudes as illustrated in Sec.~\ref{spec}. The corresponding methodology to determine the eigenvalues is described in Sec.~\ref{sec_IIB}. The eigenvalues representing complex wavenumber $\alpha=\alpha_r+i\alpha_i$ for a particular combination of Reynolds number $Re$, Mach number $M$, frequency $\omega_r$ and spanwise wavenumber $\beta$ are found by numerically integrating auxiliary compound matrix equations from free stream to wall ($y=0$) and satisfying corresponding dispersion relation. Spatial instability is determined when $\alpha_i<0$.  

We consider the Mach number of the flow ranging from $0.1$ (subsonic regime) to $6$ (supersonic regime) in the analysis presented here. We first validate the methodology by comparing the spatial stability of the compressible $M=0.1$ boundary layer with the corresponding incompressible Blasius shear layer. The variation of the spectral amplitudes for the latter is given by the Orr-Sommerfeld equation (OSE). We show an excellent match between the spatial stability characteristics for the primary and secondary modes corresponding to both flows. Subsequently, we analyze the effects of flow compressibility (Mach number $M$) and effects of spanwise wavenumber $\beta$. We note that flow becomes increasingly stable with an increase in $M$ up to $M=2$. When considering $3D$ disturbances on the $2D$ compressible mean flow, we also note that for subsonic cases, flow is increasingly stable with an increase in spanwise wavenumber $\beta$, similar to the incompressible Blasius boundary layer. In contrast, there exists an optimum value for $\beta$ up to which spatial stability reduces with an increase in $\beta$ for supersonic cases up to $M=2$. Subsequent increase in $\beta$ enhances flow stability for these cases. We also find that spatial stability of $2D$ flows to pure $2D$ and $3D$ disturbance with no spanwise variation (\textit{i.e.}, $\beta=0$) are not identical. While the latter indicates the perturbation wall-normal vorticity $\hat{\xi}_y$ and spanwise velocity $\hat{w}$ are trivially zero, the former assumes that no perturbation quantities are varying along spanwise directions as illustrated in Sec.~\ref{sec_beta_effect}. While the former is more spatially unstable up to $M\simeq 1.4$, we note the opposite scenario for $1.4<M<2$. Only one primary unstable mode is found for these cases up to $M=2$; other modes are always spatially stable.

We next investigate the spatial stability for $M>3$ corresponding to $2D$ disturbances. Mack\cite{mack1984} reported the existence of two unstable zones for such cases, which fuse to form one unstable zone with larger frequency extent at higher Mach numbers and attributed such characteristics of supersonic flows to enhanced inviscid acoustic instability. We note a series of unstable modes for $M>3$, and the number of such modes is much more than two, as reported in Mack\cite{mack1984}. The number and the frequency extent of the unstable zone for these modes increase significantly with an increase in Mach number. For example, we note $6$, $8$ and $10$ spatially unstable modes up to streamwise wavenumber of $\alpha_r=0.4$ for $M=3$, $4$ and $6$, respectively. These modes display spatial instability beyond certain critical Reynolds number $Re_{cr}$. We calculate these spatially unstable modes to travel at a much higher speed of $0.78$ to $0.89$ times the free-stream speed than those corresponding to incompressible, subsonic, and low supersonic ($M<2$) cases. While the shape of the neutral curves for the second unstable mode for $M=4$ and $6$ bears similarity to the fused neutral curve shown in Mack\cite{mack1984} for $M=4.8$, the characteristics of spatially unstable higher-order modes, to the best of our knowledge, have not been shown or reported in any related literature so far considering viscous stability of supersonic boundary layer.

\section{Acknowledgments}
SB acknowledges the support provided by DST(SERB) under MATRICS (MTR/2020/000568) and CRG (CRG/2021/006377) schemes.

%% \bibliographystyle{elsarticle-num}
%% \bibliography{Reference}

\providecommand{\noopsort}[1]{}\providecommand{\singleletter}[1]{#1}%

\clearpage
\clearpage
\clearpage

\newpage

%% \noindent {\textbf{Appendix-I: Governing equation for wall-normal variation of spectral amplitudes of linearized perturbations}}  \label{app_1}

 \noindent {\textbf{Appendix-$I$: The formulation of the CMM}} \label{app_2}

The equation for the disturbance spectral amplitudes, given by Eqs.~(\ref{app1_eq1}-\ref{app1_eq5}), are cast in the form of a system of first-order ODEs following Eq.~(\ref{eq15}), where the state vector $\left\{\textbf{X}\right\}=\left[X_1,X_2,X_3,X_4,X_5,X_6,X_7,X_8\right]^T$ are defined as $X_1=\psi$, $X_2=\psi^{\prime}$, $X_3=\varphi$, $X_4=\varphi^{\prime}$, $X_5=\Theta$, $X_6=\Theta^{\prime}$, $X_7=\Omega$ and $X_8=\Omega^{\prime}$ for $3D$ disturbances. The elements of the matrix $[E]$ are obtained from Eqs.~(\ref{app1_eq1}-\ref{app1_eq5}). We can express Eq.~(\ref{eq15}) using tensorial index notation as

 \begin{eqnarray}
   X_j=E_{jk}X_k \label{app2_eq1a}
\end{eqnarray}
 
\noindent There are four modes decaying in the free-stream for $3D$ disturbance, and one can express the state vector as a linear combination of the four decaying physical modes, following Eq.~(\ref{eq12a}) as

\begin{eqnarray}
   X_j^{\prime} = c_1 X_{j,1}+c_3X_{j,3}+c_5X_{j,5}+c_7X_{j,7}
   \label{app2_eq1}
\end{eqnarray}    

\noindent where $j=1,\cdots,8$. The auxiliary compound variables for CMM are defined following Eq.~(\ref{eq16a}). While defining the compound variables, we need to maintain the following order among the indices as $1\le j < k < l < m \le 8$, so that there is no repetition of the basis row-vectors $h_j=[X_{j,1},X_{j,3},X_{j,5},X_{j,7}]$. We have denoted the compound variables as $Y_n=\mathcal{Y}_{j,k,l,m}=\det\left([h_j,h_k,h_l,h_m]^T\right)$, where $\det(\cdot)$ represents the determinant of a square matrix. For given $j$, $k$, $l$, and $m$ indices corresponding to the state vector $X$, the index $n$ of the compound variable $Y_n$ can be obtained by sequentially counting it such that $1\le j < k < l < m \le 8$. Following this methodology, we can mathematically express $n$ as a function of $j$, $k$, $l$, and $m$ indices as
\begin{eqnarray}
  n & = & \sum_{j_1=1}^{j-1} \sum_{k_1=j_1+1}^{6} \sum_{l_1=k_1+1}^{7} \sum_{m_1=l_1+1}^{8} \left(1\right) + \sum_{k_1=j+1}^{k-1} \sum_{l_1=k_1+1}^{7} \sum_{m_1=l_1+1}^{8} \left(1\right) \nonumber \\
   & & + \sum_{l_1=k+1}^{l-1} \sum_{m_1=l_1+1}^{8} \left(1\right) + \sum_{m_1=l+1}^{m-1} \left(1\right) +1  \label{app2_eq2} 
\end{eqnarray}

\noindent The compound variable index $n$ by evaluating the above series is given as $n=(m-l)+(l-k-1)(16-l-k)/2+(k-j-1)(168-23k-22j+k^2+j^2+kj)/6+(j-1)(-j^3+29j^2-306j+1344)/24$.
The inverse calculation, \textit{i.e.}, for a given index $n$ of the compound variable $Y_n$, the indices $j$, $k$, $l$, and $m$ of the constituent basis row-vectors can also be obtained from the above algorithm or the formulae given in Eq.~(\ref{app2_eq2}).  

\noindent {\textbf{Appendix-$II$: Derivatives of the compound variable}} \label{app_3} 

Here, we explain the methodology involved in finding the first derivative of the $n^{th}$ compound variable $Y_n$ so that elements of $[F]$-matrix (see Eq.~(\ref{eq17})) can be evaluated. It is quite easy to explicitly evaluate the coefficients of the $[F]$ matrix for the Orr-Sommerfeld equation\cite{drazin2004}, primarily because the $[E]$ matrix which relates $\textbf{X}^{\prime}$ with $\textbf{X}$ (see Eq.~(\ref{eq15})) is significantly sparse in nature. For the present case, the $[E]$ matrix is relatively dense for $6^{th}$- or $8^{th}$-order cases corresponding to $2D$ and $3D$ disturbances. Therefore, we develop an algorithm based on several well-known theorems on the properties of the determinant to circumvent the manual calculation of the coefficients of the $[F]$ matrix. The proposed algorithm is generic and can be used for even higher-order systems if one considers chemical reactions, non-equilibrium thermodynamics, real-gas effects, or electro-magnetodynamic effects. This algorithm is illustrated next.

First, we enumerate several well-known theorems on determinants that are used to derive the auxiliary equations following CMM. Let $\det(\textbf{A})$ represents the determinant of $n\times n$ square matrix $\textbf{A}=[\textbf{a}_1,\textbf{a}_2,\cdots,\textbf{a}_i,\cdots,\textbf{a}_n]^T$, where, $\textbf{a}_i$ represents the $i^{th}$ row of $\textbf{A}$.  

\begin{enumerate}
\item The derivative of $\det(\textbf{A})$ is equal to the sum of $n$ auxiliary determinants of $n\times n$ square matrices $\textbf{A}_i$ such that the matrix $\textbf{A}_i$ is equal to $\textbf{A}$ except for the $i^{th}$-row, where the elements of $\textbf{A}$ are replaced by their respective derivatives. Mathematically, $\det(\textbf{A})^{\prime}=\sum_{i=1}^n \det(\textbf{A}_i)$, where $\textbf{A}_i=\left([\textbf{a}_1,\textbf{a}_2,\cdots,\textbf{a}_i^{\prime},\cdots,\textbf{a}_n]^T\right)$.
  
\item If two rows of the matrix $\textbf{A}$ are interchanged, the value of the determinant is multiplied by the factor of $(-1)$,
  \textit{i.e.}, \newline $\det\left([\textbf{a}_1,\textbf{a}_2,\cdots,\textbf{a}_i,\cdots,\textbf{a}_j,\cdots,\textbf{a}_n]^T\right)=-\det\left([\textbf{a}_1,\textbf{a}_2,\cdots,\textbf{a}_j,\cdots,\textbf{a}_i,\cdots,\textbf{a}_n]^T\right)$.

\item If the $i^{th}$-row of $\textbf{A}$ is multiplied by a constant scalar $c$, then $\det(\textbf{A})$ is multiplied by the same scalar $c$,
  \textit{i.e.}, $\det\left([\textbf{a}_1,\textbf{a}_2,\cdots,c\textbf{a}_i,\cdots,\textbf{a}_n]^T\right)=c\cdot \det\left([\textbf{a}_1,\textbf{a}_2,\cdots,\textbf{a}_i,\cdots,\textbf{a}_n]^T\right)=c\cdot \det(\textbf{A})$.

\item The determinant is \textit{multilinear}, \textit{i.e.}, if the $i^{th}$-row of $\textbf{A}$ is written as a linear combination as $\textbf{a}_i=c_1\textbf{u}+c_2\textbf{v}$, where $\textbf{u}$ and $\textbf{v}$ are row vectors and $c_1$ and $c_2$ are scalars, then $\det(\textbf{A})$ can be expressied as a similar linear combination of two determinants as $\det(\textbf{A})=\det\left([\textbf{a}_1,\textbf{a}_2,\cdots,\textbf{a}_i,\cdots,\textbf{a}_n]^T\right)=\det\left([\textbf{a}_1,\textbf{a}_2,\cdots,c_1\textbf{u}+c_2\textbf{v},\cdots,\textbf{a}_n]^T\right)=c_1\cdot\det\left([\textbf{a}_1,\textbf{a}_2,\cdots,\textbf{u},\cdots,\textbf{a}_n]^T\right)+c_2\cdot\det\left([\textbf{a}_1,\textbf{a}_2,\cdots,\textbf{v},\cdots,\textbf{a}_n]^T\right)$.

\item If two rows of $\textbf{A}$ are identical, its determinant is zero, \textit{i.e.}, \newline
  $\det\left([\textbf{a}_1,\textbf{a}_2,\cdots,\textbf{u},\cdots,\textbf{u},\cdots,\textbf{a}_n]^T\right)=0$.  
\end{enumerate}   

\noindent We can determine wall-normal derivative of $Y_n$ using these theorems and fact that $X_{j,q}^{\prime}=\sum_{l=1}^{8}E_{jl}X_{l,q}$, where $q=1,3,5,7$ represents the number corresponding to decaying physical modes in the free-stream for the $8^{th}$-order system. Let, $h_j=[X_{j,1},X_{j,3},X_{j,5},X_{j,7}]$ where $j=1,\cdots,8$ and hence, $Y_{n}=\mathcal{Y}_{j,k,l,m}=\det\left([h_j,h_k,h_l,h_m]^T\right)$, where $1\le j<k<l<m\le 8$ following Eq.~(\ref{eq16a}). Therefore,  
\begin{eqnarray}
  Y_{n}^{\prime}=\mathcal{Y}_{j,k,l,m}^{\prime}=
  \underbrace{
    \begin{vmatrix}
    h_j^{\prime} \\
    h_k \\
    h_l \\
    h_m
    \end{vmatrix}
  }_{\text{Term-1}}
  +
  \underbrace{
    \begin{vmatrix}
    h_j \\
    h_k^{\prime} \\
    h_l \\
    h_m
    \end{vmatrix}
    }_{\text{Term-2}}
  +
  \underbrace{
  \begin{vmatrix}
    h_j \\
    h_k \\
    h_l^{\prime} \\
    h_m
  \end{vmatrix}
  }_{\text{Term-3}}
  +
  \underbrace{
  \begin{vmatrix}
    h_j \\
    h_k \\
    h_l \\
    h_m^{\prime}
  \end{vmatrix}
  }_{\text{Term-4}}
  \label{app2_eq3}
\end{eqnarray}

\noindent We also know $h_j^{\prime}=E_{jp} h_p$, following Eq.~(\ref{eq15}). Therefore, we can express terms $1-4$ of Eq.~(\ref{app2_eq3}) by putting the above relation into it, maintaining the order of the indices to define $Y_n$ and using theorems $1-5$ on the determinants as 
\begin{eqnarray}
%%   % ------------ Term-1
%%   \begin{vmatrix}
%%     h_j^{\prime} \\
%%     h_k \\
%%     h_l \\
%%     h_m
%%   \end{vmatrix}
  \text{Term-1} & = & \left( 
   \sum_{p=1  }^{k-1} E_{jp} \mathcal{Y}_{p,k,l,m}
  -\sum_{p=k+1}^{l-1} E_{jp} \mathcal{Y}_{k,p,l,m}
  +\sum_{p=l+1}^{m-1} E_{jp} \mathcal{Y}_{k,l,p,m}
  -\sum_{p=m+1}^{8  } E_{jp} \mathcal{Y}_{k,l,m,p}
  \right) \hspace{1.0cm} \label{app2_eq4a} \\
%% % --------- Term-2 
%%   \begin{vmatrix}
%%     h_j \\
%%     h_k^{\prime} \\
%%     h_l \\
%%     h_m
%%   \end{vmatrix}
  \text{Term-2} & = & \left( 
  -\sum_{p=1  }^{j-1} E_{kp} \mathcal{Y}_{p,j,l,m}
  +\sum_{p=j+1}^{l-1} E_{kp} \mathcal{Y}_{j,p,l,m}
  -\sum_{p=l+1}^{m-1} E_{kp} \mathcal{Y}_{j,l,p,m}
  +\sum_{p=m+1}^{8  } E_{kp} \mathcal{Y}_{j,l,m,p}
  \right) \hspace{1.0cm} \label{app2_eq4b} \\
%%     % ------------ Term-3 
%%   \begin{vmatrix}
%%     h_j \\
%%     h_k \\
%%     h_l^{\prime} \\
%%     h_m
%%   \end{vmatrix}
  \text{Term-3} & = & \left( 
   \sum_{p=1  }^{j-1} E_{lp} \mathcal{Y}_{p,j,k,m}
  -\sum_{p=j+1}^{k-1} E_{lp} \mathcal{Y}_{j,p,k,m}
  +\sum_{p=k+1}^{m-1} E_{lp} \mathcal{Y}_{j,k,p,m}
  -\sum_{p=m+1}^{8  } E_{lp} \mathcal{Y}_{j,k,m,p}
  \right) \hspace{1.0cm} \label{app2_eq4c} \\
%%   % -------------- Term-4
%%   \begin{vmatrix}
%%     h_j \\
%%     h_k \\
%%     h_l \\
%%     h_m^{\prime}
%%   \end{vmatrix}
  \text{Term-4} & = & \left( 
  -\sum_{p=1  }^{j-1} E_{lp} \mathcal{Y}_{p,j,k,l}
  +\sum_{p=j+1}^{k-1} E_{lp} \mathcal{Y}_{j,p,k,l}
  -\sum_{p=k+1}^{l-1} E_{lp} \mathcal{Y}_{j,k,p,l}
  +\sum_{p=l+1}^{8  } E_{lp} \mathcal{Y}_{j,k,l,p}
  \right) \hspace{1.0cm} \label{app2_eq4d}
\end{eqnarray}

\noindent The final expression for $Y_n^{\prime}$ is obtained by adding Eqs.~(\ref{app2_eq4a}-\ref{app2_eq4d}). These linear ODEs need to be integrated from the free-stream to the wall. Therefore, the condition at the free-stream acts as an initial value for these equations.

\end{document}